%% file: twisted.tex
\documentstyle[
12pt]{amsart}

\newtheorem{thm}{Theorem}[section]
\newtheorem{cor}[thm]{Corollary}
\newtheorem{lem}[thm]{Lemma}

\theoremstyle{definition}
\newtheorem{defn}[thm]{Definition}

\theoremstyle{remark}

\numberwithin{equation}{section}

\newcommand{\thmref}[1]{Theorem~\ref{#1}}

\newcommand{\lemref}[1]{Lemma~\ref{#1}}

\newcommand{\bg}{\begin}

\newcommand{\no}{\noindent}

\begin{document}
\include{title}

\include{sect1}

\include{sect2}

\include{sect3}

\include{sect4}

\include{sect5}

\include{sect6}

\include{sect7}
\include{biblio}

\include{figures}

\end{document}

%% file: title.tex
\vskip .4in

\begin{center} {\bf Twisted and Nontwisted Bifurcations Induced by Diffusion}
\end{center}
\begin{center}  Xiao-Biao Lin\footnote{Department of Mathematics, North
Carolina State University, Raleigh, N.C. 27695-8205.}
\end{center}
\vskip .2in

\noindent {\bf Abstract}:\quad We discuss a diffusively perturbed predator-prey
system. Freedman and Wolkowicz showed that the corresponding ODE can have a
periodic solution that bifurcates from a homoclinic loop. When the diffusion
coefficients are large, this solution represents a stable, spatially homogeneous
time-periodic solution of the PDE. We show that when the diffusion coefficients
become small, the spatially homogeneous periodic solution becomes unstable and
bifurcates into spatially nonhomogeneous periodic solutions. The nature of the
bifurcation is determined by the twistedness of an equilibrium/homoclinic
bifurcation that occurs as the diffusion coefficients decrease. In the
nontwisted case two spatially nonhomogeneous simple periodic solutions of equal
period are generated, while in the twisted case a unique spatially
nonhomogeneous double periodic solution is generated through period-doubling.

\vskip 2.2in

%% file: sect1.tex
\section{Introduction}

Suppose that the ODE system
\begin{equation}\label{E1.1}
U'=F(U,k), \quad U\in{\Bbb R}^{2},\,k\in{\Bbb R},\, F\in C^\infty({\Bbb
R}^2\times {\Bbb R} ;
{\Bbb R}^2),
\end{equation}
has a homoclinic solution \(U=q(t)\) when the parameter \(k=k_{\infty }.\)
Assume also that for $k_\infty-\epsilon<k<k_\infty$, there is a stable periodic 
solution $ U=p(t,k) $ bifurcating from $q(t)$.
We study the diffusively perturbed system
\begin{equation}\label{E1.2} \begin{array}{ll}
U_{t}=DU_{\xi \xi }+F(U,k), & \quad 0< \xi < 1, \\[8pt] U_{\xi }(t,0 )=
U_{\xi }(t,1 )=0, \end{array}
\end{equation}
where $D=\text{diag}\{d_{1},d_{2}\}$ is a positive diagonal matrix. The boundary
conditions ensure that $ U(t,\xi )=q(t) $ or $ U(t,\xi )=p(t,k) $ is still a
solution for system (\ref{E1.2}). Results from  \cite{conwayhoffsomller,hale86}
indicate that when the diffusion coefficients are large, these spatially
homogeneous solutions (to be called SH solutions in the sequel) are stable.
However, when the diffusion coefficients become small, SH solutions may lose
stability and bifurcate into spatially nonhomogeneous solutions
(to be called SN solutions).

Such a bifurcation can create spatially nonhomogeneous
patterns. Existing literature on pattern
generation concentrates on small patterns generated through bifurcations of
equilibria, or traveling waves constructed using transition layers 
\cite{turing52,conway}. The 
mechanism of pattern generation studied in this paper is fundamentally 
different.

Since many ODE models are approximations to more realistic models where
diffusion is present, examples that lead to systems (\ref{E1.1}) and
(\ref{E1.2}) are plentiful.
Freedman and Wolkowicz, \cite{freedmanwolkowicz,wolkowicz88}, studied a
two species predator-prey system that models the group defense of prey against
predation. They found a homoclinic solution $ q(t) $ at a certain parameter 
value
$ k=k_{\infty }. $ The homoclinic solution bifurcates into a long period
solution $ p(t,k) $ when $ k_{\infty }-\epsilon<k<k_{\infty }$. Suppose now 
diffusion is added to the system as in (\ref{E1.2}). The
region of $(d_{1}, d_{2})$ where $p(t,k)$ is stable has been studied
in \cite{lin92},  but
the bifurcation when parameters cross the boundary $\Gamma$ of the
region has not been discussed. The purpose of the present paper is to discuss
the bifurcation of $ p(t,k) $ when $ (d_{1}, d_{2}) $ crosses $\Gamma$. The
bifurcation of $q(t)$ into SN homoclinic solutions will be presented as the
limit when the period is infinity.

The creation of SN periodic solutions is caused jointly by the homoclinic
bifurcation in (\ref{E1.1}) and an equilibrium
bifurcation in PDE modes in (\ref{E1.2}). 
Let $k=k_\infty$ such that (\ref{E1.1}) possesses
a homoclinic solution $q(t)$ asymptotic to a hyperbolic equilibrium $E$. 
It is easy to find a curve $\Gamma$ in the
$(d_1,d_2)$--plane  where the linearization
of (\ref{E1.2}) at E has a simple zero eigenvalue, and no other eigenvalue
on the imaginary axis. When $(d_1,d_2)$ crosses $\Gamma$ transversely, the
equilibrium $E$ of (\ref{E1.2}) loses the hyperbolicity and two SN equilibria 
bifurcate from it.
To describe bifurcations of $q(t)$ and $p(t,k)$ when
$ (d_{1}, d_{2}) $ crosses $\Gamma$, we need the concept of the
twistedness of the homoclinic solution $q(t)$. Let $\phi_{c}$
be a unit eigenvector corresponding to the zero eigenvalue, unique up to the
multiplication by $-1$. It can be shown that the linearization of (\ref{E1.2})
around $q(t)$ has a
solution $\phi (t) $ that approaches $\phi_{c}$ as $t \rightarrow -\infty, $
and approaches $c^{*} \phi_{c} $ as $ t \rightarrow +\infty .$ Here $c^{*}$
is a scalar function of $(d_{1}, d_{2})$. The limit of the solution $\phi (t)$
as $t \to -\infty$ is in fact
a tangent vector to $W^{cu}_{loc}(E)$, transverse to the unstable eigenvector.
See Fig. 1.1.
We say that the homoclinic solution of (\ref{E1.2}) is twisted if $c^{*}<0$,
nontwisted
if $c^{*}>0$ and degenerate if $c^{*}=0$. 

All the three cases have been found
in Freedman and Wolkowicz's example by the numerical computation of $c^{*}$.
See Figure 7.4.  An equivalent definition of twistedness will be given in
\S3. I have recently found that the change of twistedness is a generic
phenomenon when diffusions are added to an ODE system that possesses a stable
homoclinic orbit. But the proof will require a separate paper.
\vskip1.0in
\centerline{\bf Fig. 1.1}
\vskip 0.05in

The bifurcation of  $p(t,k)$ is determined by the twistedness of 
the homoclinic solution $q(t)$ at the point where $(d_1,d_2)$ crosses $\Gamma$.
Roughly speaking, two SN simple periodic solutions of equal periods are
generated in the nontwisted case, while in the twisted case a unique SN 
symmetric double
periodic solution is generated through period-doubling. A periodic solution
$U(t,\xi)$ is said to be a simple periodic solution if its trajectory in a 
function space 
stays near the orbit of $q(t)$ and hits a cross section $\Sigma$ to the orbit
of $q(t)$ precisely once.
It is said to be a symmetric double periodic solution if it hits $\Sigma$ 
precisely two
times and satisfies a symmetry condition $U(t+T,\xi)=U(t,1-\xi)$. Here $2T$ is
the period of $U(t,\xi)$. Finally, when $q(t)$ is degenerate, it is possible to
cross $\Gamma $ in a way that no SN simple or symmetric double periodic 
solutions are generated. We do not discuss the existence of other
types of SN solutions in this
paper due to technical complications.  The homoclinic twist bifurcation at a
hyperbolic equilibrium was
discovered in \cite{yanagida87} and later studied in \cite{kokubu88} and
\cite{chowdengfiedler}. But the homoclinic twist
bifurcation discussed in this paper is new even in the ODE context.

In a separate paper, \cite{lin94}, we show how our method can be used to prove
the stability of the SN periodic solutions.

System (\ref{E1.2}) will be studied in the intermediate spaces $ D_{A}(\theta)$ and
$D_{A}(\theta +1) $, $0< \theta <1$. These function
spaces allow solutions of (\ref{E1.2}) to have so-called maximal regularity,
and are normally used to study fully
nonlinear parabolic equations \cite{dapratogrisvard}. Our system is not fully
nonlinear, but to prove the smooth dependence of the
solutions on $d_1$ and $ d_2$, which are the coefficients of the highest
derivatives
in the equations, we need to use the maximal regularity of the solutions.

Solutions of (\ref{E1.2}) satisfy a reflection symmetry about the midpoint
of the domain [0,1],
due to the special boundary conditions imposed there. For a function $U(\xi )$
defined on $\xi \in [0,1]$, let $ (RU)( \xi ) = U(1- \xi ), \: 0
\leq \xi \leq 1$. It can be verified if $U(t, \xi)$ is a solution to
(\ref{E1.2}), so is $RU(t, \xi)$. 
Consequently, if $U_1$ is a SN periodic solution and is mutually disjoint with
$RU_1$, then we have a pair of SN periodic solutions related by the symmetry.
On the other hand, if $U_1$ has a nonempty intersection with $RU_1$, then
$U_1$ is a $2T$ period SN solution 
satisfying $U(t+T , \xi) = RU(t, \xi)$. The R symmetry is very important in 
this paper since we can
show that local center manifolds and flows on them respect the symmetry group R.
We can also show that bifurcation functions derived by Lyapunov-Schmidt
procedures are invariant with respect to  R. A mapping
$f:C[0,1]\rightarrow C[0,1]$ is invariant with respect to R if $f(RU) = Rf(U)$.

Suppose now the Neumann boundary conditions in (\ref{E1.2}) is replaced by 
periodic boundary conditions. In addition
to the reflection symmetry, there is also a rotation
symmetry, i.e., $U(t, \xi + \theta)$ is a solution if $U(t, \xi)$ is a
solution. The bifurcation picture is quite different.
Spatially nonhomogeneous tori may be generated instead of periodic
solutions. See \cite{lin94}.
Periodic boundary conditions will not be pursued further in this paper.

We introduce intermediate spaces $D_{A}(\theta)$ and $D_{A}(\theta + 1)$ in \S
2. We then study invariant manifolds and their foliations in these
spaces. These invariant manifolds and their foliations provide convenient
coordinates to study dynamics of (\ref{E1.2}) near a equilibrium $E$. Some
important lemmas regarding the symmetry R are also presented there.

The assumptions and the main results of this paper are given in \S 3.

In \S4 we prove some lemmas needed in the sequel. In \S 5, we use a
Lyapunov-Schmidt type reduction to obtain a one-dimensional bifurcation 
equation whose solutions correspond to simple or 
symmetric double SN periodic solutions. Proofs of the main theorems are given 
in \S 6. In \S 7 we summarize our numerical results about the example from
\cite{freedmanwolkowicz}.

Recently Sandstede \cite{sandstede} has constructed center
manifolds around some homoclinic solutions. It is hoped that
a center manifold that is tangent to $q'(t)$ and $\phi(t)$ can be constructed
some day. And it may bifurcate into a center manifold around the orbit of
$p(t,k)$.   Thus the twistedness of the homoclinic
orbit should be passed to the twistedness of the center manifold of the periodic
orbit.
And the bifurcation of the periodic solutions should be determined by an 
one-dimensional
return map on this center manifold. Thus, we naturally expect to see the 
occurrence
of simple or symmetric double periodic solution on this center manifold.
However, I was unable to use the center manifold technique in this
paper due to technical complications. On the contrary, the bifurcation function
approach in this paper is easy
to use. The trade-off is that only simple periodic solutions
and symmetric double periodic solutions are discussed in this paper.
Complete understanding of the dynamics near the homoclinic orbit is still open,
especially around the degenerate point $c^*=0$.

%% file: sect2.tex
\section{ Abstract parabolic equations, invariant manifolds and foliations}

The PDE system (\ref{E1.2}) is studied in the intermediate spaces
$D_A(\theta)$ and $D_A(\theta+1)$.  Let $A$ be a densely defined
sectorial operator that generates a
$C_0$ analytic semigroup $e^{At}$ in a Banach space $\cal X$.  For each $0
< \theta <1$, define Banach spaces
$$
\begin{array}{ll}
D_A(\theta) = \{x \in  {\cal X} \mid \lim _{t\rightarrow
0} t^{1-\theta} Ae^{At} x = 0\},\\[8pt]
D_A(\theta + 1)  =  \{x \in D_A \mid A x \in D_A(\theta)\}.
\end{array}
$$
The norms are
$$
\begin{array}{ll}
\|x\|_\theta  =   \sup _{0< t \leq 1} \|t^{1-\theta} Ae^{At} x \|_ {\cal X} +
\|x\|_{\cal X},\\[8pt]
\|x\|_{\theta +1}  =  \|Ax\|_\theta + \|x\|_{\cal X}.
\end{array}
$$
Intermediate spaces $D_A(\theta + m),\ 0 < \theta < 1,\ m\in {\Bbb N}^+$ can be
defined similarly.  Throughout this paper, let ${\cal X} = [L^2(0,1)]^2,\
A =\left ( \begin{array}{lr} \partial_{\xi \xi}& {}  \\
{} & \partial _{\xi\xi}\end{array}\right )$
and $D_A =  \{u\in$ $[H^2(0,1)]^2, u_\xi(0) = u_\xi(1) = 0\}$.
See \cite{dapratogrisvard} for details about the intermediate spaces.

Denote ${\cal F}(U,d_1,d_2,k) = DU_{\xi\xi} + F(U,k)$.  We can write (\ref{E1.2}) as an
abstract nonlinear parabolic equation
\begin{equation}\label{E2.1}
U' = {\cal F}(U,\mu)
\end{equation}
where $\mu = (d_1,d_2,k)$ is a parameter. The solution for (\ref{E2.1}) with
$U(0) = U_0$ will be denoted by $U_*(t,U_0)$.

It is easy to see that ${\cal F}:D_A(\theta +m+1)\times{\Bbb R}^3
\rightarrow D_A(\theta +m),\,m\geq 0$ is $C^\infty$. Also
$F:D_A\to\cal{X}$ is $C^\infty$ since $F$ is $C^\infty$.  The
following existence theorem is from \cite{dapratogrisvard}.

\begin{thm}\label{T2.1}
For each $U_0\in D_A(\theta + 1)$, there exists
$\tau > 0$ so that (\ref{E2.1}) admits a unique solution $U \in C^1([0,\tau];
D_A(\theta)) \cap C^0 ([0,\tau]; D_A(\theta + 1))$.  Moreover, $U$ is a
$C^\infty$ function of $U_0$ and $\mu$ in the specified function spaces.
\end{thm}

Consider a time dependent linear system
\begin{equation}\label{E2.2}
\begin{array}{lll}u'& =& A(t) u + f(t),\\
u(s) &=& x,\quad a \leq s \leq t \leq b,\end{array}
\end{equation}
which comes from linearizing (\ref{E2.1}) around a particular
solution. It is easy to verify that

\noindent (1) for all $t \in [a,b],\, A(t) : D_{A(t)} \rightarrow \cal{X}$ is
sectorial, $D_{A(t)} = D_A$ with equivalent norms;

\noindent (2) for each $0 < \theta < 1$,
$$
D_{A(t)} (\theta +1) = D_{A(\tau)} (\theta + 1) = D_A(\theta + 1),
\hbox { for all } t, \tau \in [a,b] , $$
with equivalent norms;

\noindent (3) $A(\cdot ) \in C([a,b]; L(D_A,{\cal X})) \cap C([a,b];
L(D_A(\theta + 1), D_A(\theta))).$

\noindent The following theorem is proved in  [1].

\begin{thm}\label{T2.2}  Under the above conditions, there is a unique
solution
$$
U \in C([s,b]; D_A(\theta + 1)) \cap C^1 ([s,b]; D_A(\theta))
$$
to  (\ref{E2.2}) for each $x \in D_A(\theta + 1)$ and $f
\in C([s,b]; D_A(\theta))$.  Denote the solution by $U(t) = T(t,s)x$ when $f =
0$.  Then $T(t,s)$ extends to $D_A(\theta)$ by continuity.  Finally, the
variation of constants formula holds for solutions of (\ref{E2.2}) with
$f \neq 0$:
$$
U(t) = T(t,s) x + \int^t_s T(t,\xi) f(\xi) d\xi.
$$
\end{thm}

Using the variation of constants formula, many familiar results of ODE systems
can be extended to (\ref{E2.1}), with almost identical proofs.  The most useful ones in this paper are the smoothness of invariant manifolds and their
foliations.

After a shift of coordinates, assume that $\{0\} \in D_A(\theta +1)$ is an
equilibrium of (\ref{E2.1}).  Let $\tilde A= D_U{\cal F}(0,\,\mu_0)$ where
$D_{\tilde A}=D_A$ and $\mu_0=(d_{10},d_{20},k_\infty)$.
Here $(d_{10},d_{20})\in\Gamma$ so that zero is an eigenvalue for $\tilde A$.
Let ${\cal N}={\cal F}- \tilde A U$.  System (\ref{E2.1}) can be written as
$$ U'=\tilde AU+{\cal N}(U,\mu).$$
Let
$$
\begin{array} {lll}
\sigma({\tilde A}) &= &\sigma_-  \cup \sigma_0 \cup \sigma _+\\
Re\,\sigma_- &\leq &- \lambda _M,\\
Re\, \sigma _+ &\geq &\lambda _M,\\
Re\, \sigma _0& =& 0.
\end{array}
$$
for some $\lambda_M  > 0$.  Let $X, Y$ and $Z$ be the invariant subspaces
corresponding to the spectral set $\sigma _+, \sigma _-\mbox{ and }\sigma _0$
respectively.  We will identify $D_A(\theta +1)$ with $X \times Y \times Z$ by
writing $ U = (x,y,z)$ if $U = x+y+z$. Since $X,\,Y,\,Z\subset D_A(\theta+1)$,
$Rx$,  $Ry$ and $Rz$ are defined by
restricting $R$ to  subsets of $D_A(\theta+1)$. It is also easy to verify
that $RX=X,\,RY=Y$ and $RZ=Z$.

Since both ${\cal N}:D_A(\theta +1) \times {\Bbb R}^3
\rightarrow D_A(\theta)$ and ${\cal N}: D_A \times {\Bbb R}^3
\rightarrow {\cal X}$ are $C^\infty$, there exists a local center manifold that
is $C^\nu$ for any $\nu>0$, \cite{lunardi87,chowlu88}.  Using the
method of \cite{chowlinlu91}, which treats semi-linear parabolic equations, we
can prove that the center unstable and center stable manifolds
are $C^\nu$, and there exists
$C^{\nu}$ invariant foliation of center unstable (center stable) manifolds by
unstable (stable) fibers, if Lip $\cal N$ is small.
The smallness of Lip $\cal N$ can be
removed by modifying the equation outside a neighborhood of $\{0\}$, if we
are only interested in local invariant manifolds and their foliations.
In the following we show how to choose the modifier so that the reflection
symmetry resulting from the Neumann boundary conditions will be preserved for
the induced flow on the local center manifold.

First we give coordinate-free definitions of global center unstable, center
stable manifolds and their foliations.  See \cite{chowlinlu91} for similar
definitions given to semilinear systems.

\begin{defn}\label{D2.3}  Let $0 < \lambda _1 < \lambda _2 < \lambda _M$.
The global center stable manifold is defined by
$$
\begin{array}{lcr}
W^{cs} &=& \{U_0 \in D_A(\theta + 1) \mid U_* (t,U_0) \hbox { exists for all } t
\geq 0 \hbox { and }\\[8pt]
& &\|U_*(t,U_0)\|_{\theta +1} \leq C e^{t\lambda _1},\ t \geq 0\}.
\end {array}
$$
The global center unstable manifold is defined by
$$\begin{array}{rcr}
W^{cu} &=& \{U_0 \in D_A(\theta + 1) \mid U_* (t,U_0) \hbox { exists for all } t
\leq 0 \hbox { and }\\
& &\|U_*(t,U_0)\|_{\theta +1} \leq C e^{-t\lambda_1},\ t \leq 0\}.
\end{array}$$
Define the global center manifold by
$$
W^c = W^{cu} \cap W^{cs}.
$$
For each $U_0 \in W^{cs}$ (or $W^{cu}$), the stable fiber $W^s(U_0)$ (or
unstable fiber $W^u(U_0))$ passing through $U_0$ is defined by
$$\begin{array}{rcl}
W^s(U_0)&=&\{V_0\in W^{cs}(0)\mid \|U_*(t,U_0)-U_*(t,V_0)\|_{\theta+1}\leq
Ce^{-t\lambda_2},\;t\geq0\},\\[8pt]
W^u(U_0)&=&\{V_0\in W^{cu}(0)\mid \|U_*(t,U_0)-U_*(t,V_0)\|_{\theta+1}\leq
Ce^{t\lambda_2},\;t\leq0\}.
\end{array}
$$
\end{defn}

Obviously, $W^{cs}$ is forward invariant and $W^{cu}$ is backward invariant.
$W^c$ is invariant. Also each point on $W^{cs}$ (or $W^{cu}$) belongs to one
and only one stable (or unstable) fiber. The global foliations are forward or
backward invariant in the sense that
$$
\begin{array}{lll}
U_*(t,W^s(U_0)) &\subset &W^s (U_*(t,U_0)),\quad t \geq 0;\\[6pt]
U_*(t,W^u(U_0)) &\subset &W^u (U_*(t,U_0)),\quad t \leq 0.
\end{array}
$$

Let ${\cal O}\subset D_A(\theta +1)$ be an open set containing the equilibrium
$\{0\}$.  Let $\tilde {\cal F}: D_A(\theta + 1) \times {\Bbb R}^3 \rightarrow
D_A(\theta)$ and $\tilde {\cal F}: D_A \times {\Bbb R}^3
\rightarrow \cal{X}$ be $C^\nu,\, \nu > 0$.
Assume that $\tilde {\cal F}= {\cal F}$ in ${\cal O}\times {\Bbb R}^3$.
Consider the system
\begin{equation}\label{E2.3}
U' = \tilde {\cal F}(U,\mu).
\end{equation}

\begin{defn}\label{D2.4} Assume that (\ref{E2.3}) has global invariant center
stabel and center unstable manifolds, and invariant foliations as defined in
Definition \ref{D2.3}.
Local invariant manifolds $W^{cu}_{loc}$,
$W^{cs}_{loc}$ and $W^c_{loc}$ for system (\ref{E2.1}) are the restrictions
to $\cal O$ of the global invariant manifolds for system (\ref{E2.3}).
Local invariant foliations of $W^{cs}_{loc}$ and $W^{cu}_{loc}$ for system
(\ref{E2.1}) are the restrictions to $\cal O$ of the global invariant
foliations of $W^{cu}$ and $W^{cs}$ for system (\ref{E2.3}).
\end{defn}

Local invariant manifolds and local invariant foliations  depend on the
extension of ${\cal F}$  to $\tilde {\cal F}$ outside $\cal O$ and are
thus not unique.
Observe that $Lip\;\cal N$ is small inside $\cal O$ if the neighborhood
$\cal O$ is
small, due to the fact ${\cal N}(0) = 0$ and ${\cal N}'(0) = 0$.
The purpose of extending
${\cal F}$ to $\tilde {\cal F}$ is to have a small Lipschitz number
for $\tilde {\cal N} = \tilde {\cal F} - {\tilde A}U$ outside $\cal O$.

Observe that $D_A(\theta + 1) \subset [H^2(0,1)]^2$ is a continuous injection
and $\|u\|_{H^2(0,1)}: H^2(0,1)\setminus\{0\} \rightarrow {\Bbb R^+}$ is
$C^\nu$ for any $\nu > 0$.
Let $\psi: {\Bbb R} \to {\Bbb R}$ be $C^\infty$ such that
$$   \begin{array}{l} \psi (s) = 1 \mbox { for } |s| \leq 1
\mbox { and } \psi(s) = 0 \mbox{ for } |s|\geq 2,\\
0 \leq \psi(s) \leq 1. \end{array}.
$$
Let $\tilde {\cal N}(U,\mu) = {\cal N}(\psi(|U|_{[H^2(0,1)]^2}/\rho)U,\mu)$,
where $\rho > 0$. It can be verified that 
$$
\tilde {\cal N}:[H^2(0,1)]^2\times {\Bbb R}^3\rightarrow [H^2(0,1)]^2
$$
is $C^\nu$, $Lip\tilde {\cal N} \rightarrow 0$ as $\rho
\rightarrow 0$, and $\tilde {\cal N}
= {\cal N}$ for $\|U\|_{[H^2(0,1)]^2} \leq \rho$.
Recall that $D_A=\{U\in [H^2(0,1)]^2\,:\partial_x U=0\,\hbox { at }x=0,1\}$.
After checking the boundary conditions, we find that both ${\cal N},
\tilde{\cal N}:D_A\to D_A$ are $C^\nu$ for any $\nu>0$.
Since $D_A(\theta + 1) \subset D_A \subset D_A(\theta)$,
$\tilde {\cal N} : D_A(\theta + 1) \to D_A(\theta)$ is $C^\nu$ with Lip $\tilde
{\cal N} \to 0$ as $\rho \to 0$ in such space.  We can prove the
following theorem by using the method employed in \cite{chowlinlu91}.

\begin{thm}\label{T2.5} For any $\nu>0$, there exists a small constant $\rho
> 0$ such that
the global invariant manifolds for system (\ref{E2.3}) are $C^\nu$ embedded
submanifolds in $D_A(\theta +1)$ if $|\mu - \mu_0| < \rho$.  Moreover
\begin{eqnarray*}
W^{cs} & = & \{x = h_1(y,z,\mu)\},\\
W^{cu} & = &  \{y = h_2(x,z,\mu)\},
\end{eqnarray*}
where $(x,y,z) \in X \times Y \times Z$ and $|\mu - \mu_0| < \rho$.  The
function $h_i, i = 1,2$,
is $C^\nu$ in all the variables,
with $h_i(0,0,\mu_0) = 0, Dh_i(0,0,\mu_0) = 0$ and $Dh_i = O(\rho)$.

By a $C^\nu$ change of variable $(x,y,z) \to (x^1,y^1,z^1)$,
\begin{eqnarray*}
x^1 & = & x-h_1(y,z,\mu),\\
y^1 & = & y-h_2(x,z,\mu),\\
z^1  & = & z,
\end{eqnarray*}
we can flatten these manifolds,
\begin{eqnarray*}
W^{cs} &  = &  \{x^1 = 0\},\\
W^{cu} & = & \{y^1 = 0\},\\
W^c & = & \{x^1 = 0,y^1 = 0\}.
\end{eqnarray*}
The change of variables preserves the symmetry, i.e.,
$U=(x,y,z)\to (x^1,y^1,z^1)$, implies $RU=(Rx,Ry,Rz)\to (Rx^1,Ry^1,Rz^1)$.
\end{thm}

\bg{pf} The existence and smoothness of such $h_i,\, i=1,2$,
can be proved similar to \cite{chowlinlu91}.
It can be verified that $\tilde {\cal N}(Rx,Ry,Rz,\mu)=
R\tilde {\cal N}(x,y,z,\mu)$.  Thus $RW=W$, where $W$ stands for
$W^{cu}, W^{cs}$ or $W^c$.

Let $U=(x,y,z) \in W^{cu}$.  Then $RU\in W^{cu}$.  Thus $h_2(Rx,Rz,\mu) =
Ry = Rh_2(x,z,\mu)$.  Similarly, $h_1(Ry,Rz,\mu) = Rh_1(y,z,\mu)$.  It follows
that \newline $R U \to (Rx^1,Ry^1,Rz^1)$.
\end{pf}

We will use the new coordinates $(x^1,y^1,z^1)$ to discuss invariant foliations
for system (\ref{E2.3}).  Let $U_0 \in W^c$.  Then $U_0 = (0,0,z_0)$ in the new
coordinates.  Denote $W^s(U_0)$ and $W^u(U_0)$ by $W^s(z_0)$ and $W^u(z_0)$.
From Definition \ref{D2.3}, we can show that different points on $W^c$ do not  belong
to a same fiber $W^u(z_0)$ or $W^s(z_0)$. Further more, we also know that
all fibers on $W^{cu}$ (or $W^{cs}$) have to intersect $W^c$.

\begin{thm}\label{T2.6}
If $\rho > 0$ is small enough, then the stable
fibers $W^s(z_0), z_0 \in W^c$ form an invariant foliation of $W^{cs}$ and the
unstable fibers $W^u(z_0)$, $z_0 \in W^c$ form an invariant foliation of
$W^{cu}$, for $| \mu  - \mu _0 | < \rho$.  Moreover,
\begin{eqnarray*}
W^s(z_0) & = & \{ x^1 = 0, z^1 = z_0 + h_3(y^1,z_0,\mu) \mbox { with } h_3
(0,z_0,\mu) = 0\},\\
W^u(z_0) & = & \{y^1 = 0, z^1 = z_0 + h_4 (x^1,z_0,\mu) \mbox { with }
h_4(0,z_0,\mu) = 0\}.
\end{eqnarray*}
The function $h_i,\,i = 3,4$, is $C^{\nu}$ in all its variables,
$D_yh_3(0,0,\mu_0) = 0,\ D_xh_4(0,0,\mu_0) = 0$ and $Dh_i = O(\rho), i = 3,4$.
By a $C^{\nu}$ change of variables $(x^1,y^1,z^1)\to (x^2,y^2,z^2)$,
which is defined implicitly by
\begin{eqnarray*}
x^1 & = & x^2,\\
y^1 & = & y^2,\\
z^1 & = & z^2 + h_3(y^2,z^2,\mu) + h_4(x^2,z^2,\mu),
\end{eqnarray*}
we can flatten the fibers, so that
\begin{eqnarray*}
W^s(z_0) & = & \{x^2 = 0,\ z^2 = z_0\},\\
W^u(z_0) & = & \{y^2 = 0,\ z^2 = z_0\}.
\end{eqnarray*}
The change of variables preserves the symmetry,
i.e.,  if $(x^1,y^1,z^1)\to (x^2,y^2,z^2)$ then $(Rx^1,Ry^1,Rz^1)
\to (Rx^2, Ry^2,Rz^2)$.

The change of variable here does not affect the flow on $W^c$.
\end{thm}

\bg{pf}  The existence and smoothness of $h_i,\,i=3,4$, are
proved similar to that in \cite{chowlinlu91}.

i)\quad Since $ W^s(Rz_0)=RW^s(z_0)$, and $W^u(Rz_0)=RW^u(z_0)$,
we have  $h_3(Ry^1,Rz_0,\mu\\ = Rh_3(y^1,z_0,\mu)$ and
$h_4(Rx^1,Rz_0,\mu) = Rh_4(x^1,z_0,\mu)$.  It follows that $(x^1,y^1,z^1) \to
(x^2,y^2,z^2)$ implies that $(Rx^1,Ry^1,Rz^1) \to (Rx^2,Ry^2,Rz^2)$.

ii)\quad If $x^2 = 0$ and $y^2 = 0$, then $h_3(0,z^2,\mu) = 0$ and
$h_4(0,z^2,\mu) = 0$.  Therefore $z^1 = z^2$ on $W^c$.  The equation for the
flow on $W^c$ is not  changed.
\end{pf}

Define a function space $ {\cal X} _{n} = \{(u_{n}\cos( n\pi \cdot),
v_{n}\cos (n\pi \cdot)), \; (u_{n}, v_{n}) \in {\Bbb R}^{2}\} $.
Obviously ${\cal X} _{n}$ is isomorphic to
${\Bbb R}^{2}$. Observe that $\sum \{{\cal X}_{n}, \: n\geq 0\} $ is dense in
$[L^{2}(0,1)]^{2}$.

Recall that (\ref{E2.1}) comes from (\ref{E1.2}). The hypotheses on
$F$ will be specified in \S 3. In particular, they imply

1)\quad $Z$ is one dimensional, spanned by an eigenvector in ${\cal X}_1$,
corresponding to the eigenvalue $\lambda = 0$;

2) \quad $X$ is one dimensional, spanned by an eigenvector in ${\cal X}_0$,
corresponding to the eigenvalue $\lambda = \lambda _+$.

We may identify $Z$ and $X$ with ${\Bbb R}$.  More precisely, let
$w$ be a unit vector in $Z$.  For any $z \in Z$, there is a unique
$\overline z \in {\Bbb R}$ such that $z = \overline zw$.  We will identify
$z$ with $\overline z$ and drop the over-bar.  The same comment also
applies to $X$. It can be verified that if $x\in X$ and $z\in Z$, then $Rx=x$
and $Rz=-z$. We use $U \sim (x^2,y^2,z^2)$ to indicate $U$ corresponds to
$(x^2,y^2,z^2)$ in the new coordinates.

\begin{thm}\label{T2.7}  a)\quad If $U \in {\cal X}_0$ and if $U \sim
(x^2,y^2,z^2)$ in the new coordinates,  then $z^2 = 0$ and $y^2 \in {\cal
X}_0$.  The converse is also true.

\noindent b)\quad For system (\ref{E2.3}), the flow on $W^c$ has the form

\[ \begin{array}{c}
\begin{array}{cc}
x^2 =0, &  y^2=0,\end{array}\\[8pt]
\frac{d}{dt}z^2=g(z^2,\mu),
\end{array}
\]
where $g(0,\mu) = 0,\ D_{z^2}g(0,\mu_0) = 0$ and $g(-z^2,\mu) = -g(z^2,\mu)$.
\end{thm}

\bg{pf} a)\quad In the original coordinates,
$U = x+y+z$ with $x \in X$, $y \in Y$ and $z \in Z$.  If $U\in{\cal
X}_0$, then it is obvious that
$z = 0,\, x \in {\cal X}_0$ and $y \in {\cal X}_0$.

We first examine the changes of variables
$ (x,y,z)\to (x^1,y^1,z^1))$ as in \thmref{T2.5}.
Consider the change of variable $y^1 = y - h_2(x,z,\mu)$.  When $z=0$, the
graph $\{z=0,\,y = h_2(x,0,\mu)\}  = W^{cu} \cap \{z = 0\}$ is one-dimensional.  Now
let us restrict the system to ${\cal X}_0$, where $E=\{0\}$ is hyperbolic. By
the standard existence theorem of the unstable manifold for the ODE system,
there exists a smooth function $\tilde h$ such that
$W^u = \{y = \tilde h(x,\mu)\}$ for the restricted system.  Clearly the graph
$\{y = \tilde h(x,\mu)\} \subset \{y = h_2(x,0,\mu)\}$.
Since they are both one dimensional, we have that $\tilde h(x,\mu) =
h_2(x,0,\mu)$.  This proves that $h_2(x,0,\mu) \in {\cal X}_0$.
Recall that $z = z^1$.  Thus if $U \in {\cal X}_0$, $z^1 = 0$ and
$y^1 \in {\cal X}_0$, vice versa.

We now consider the second change of variable $(x^1,y^1,z^1) \to (x^2,y^2,z^2)$
as in \thmref{T2.6}. Since $y^2 = y^1$, $ y^2 \in {\cal X}_0 \Leftrightarrow
y^1 \in {\cal X}_0$.
If $(0,y^1,z^1)$ is a point on $W^s(z_0)$, then $(0, Ry^1,Rz_1)$ is a point on
$W^s(Rz_0)$. From \thmref{T2.6}, compare the $z^1$ coordinates, and observe
that $Rz_0=-z_0$, we have
$R(z_0 + h_3(y^1, z_0,\mu)) = -z_0 + h_3(Ry^1,-z_0,\mu)$. However, $Rh_3=-h_3$.
Thus $h_3(Ry^1, - z_0, \mu) = -h_3(y^1,z_0,\mu)$.  Similarly,
we can show $h_4(Rx^1,-z_0,\mu) = - h_4(x^1,z_0,\mu)$.  Therefore if
$y^2 = y^1 \in {\cal X}_0$, we have $Ry^1 = y^1$,
$h_3(y^2,0,\mu) = 0$, and $ h_4(x^2,0,\mu) = 0$.  In this case, we have
$z^1 = 0 \Leftrightarrow z^2 = 0$.

By combining the two changes of variables, we have verified the assertions of a).

\noindent b) \quad  The assertions  $x^2 = 0$ and $y^2 = 0$ are obvious.
If $U(t) \sim (0,0,z^2(t))$ is a solution on $W^c$, so is $RU(t) \sim
(0,0,-z^2(t))$. Therefore $g(-z^2,\mu) = -g(z^2,\mu)$.
\end{pf}

%% file: sect3.tex
\section { Assumptions and main results}

We assume that the ODE system (\ref{E1.1}) satisfies the following hypotheses:
\begin{itemize}
\item [$\bold H_1$)] $F:{\Bbb R}^2 \times {\Bbb R} \to {\Bbb R}^2 \hbox { is }
C^\infty$.

\item [$\bold H_2$)]  At $k = k_\infty$, (\ref{E1.1}) possesses a homoclinic solution 
$U = q(t)$ asymptotic to an equilibrium $E=E(k_\infty)$.

\item [$\bold H_3$)]  At $E(k_\infty)$, the Jacobian matrix
\end{itemize}
$$
J = \left (\begin{array}{cc}
-a & -b\\
-c & -d\end{array}\right )
$$
satisfies $a+d>0$ and $ad-bc<0$.

Hypotheses $H_3$) implies that $E(k_\infty)$ is hyperbolic with eigenvalues
denoted by $-\lambda_- < 0 < \lambda _+$, satisfying $\lambda _+ - \lambda _- <
0$.  The equilibrium $E = E(k)$ continues to exist for all $k \approx
k_\infty$.  We will suppress $k$ if no confusion should arise.  The homoclinic
orbit is stable  from inside since $\lambda_+ - \lambda _- < 0$.  Assume that
the homoclinic orbit breaks in certain direction when $k$ moves away from
$k_\infty$, so that periodic solutions bifurcate from  $q(t)$ for $k_\infty -
\epsilon < k < k_\infty$.  More precisely, consider the linear variational
equation of (\ref{E1.1}) around $U = q(t)$,
\begin{equation}\label{E3.1}
U^\prime = \partial _U F(q(t),k_\infty)U,
\end{equation}
and its adjoint equation
\begin{equation}\label{E3.2}
\Psi^\prime = - [\partial _U F(q(t), k_\infty)]^*\Psi.
\end{equation}

System (\ref{E3.2}) has a unique nontrivial bounded solution $\Psi(t)$ up to
multiplying    by nonzero constants.  It is known that $\Psi(t) \sim
\Psi_0e^{-\lambda_+t}$ and $q(-t)-E(k_\infty) \sim \phi_0e^{-\lambda_+t}$ as $t
\to + \infty$ where $\Psi _0$ (or $\phi_0)$ are left (or right) eigenvector of 
the
matrix $J$ corresponding to the eigenvalue $\lambda_+$.  See \cite{hartman64}.
For definiteness, assume
\begin{equation}\label{E3.3}
\lim_{t\to +\infty} \Psi(t) (q(-t)-E)e^{2t\lambda_+} = -1.
\end{equation}

We now assume that the breaking of the homoclinic solution $q(t)$ is in the
direction determined by

\begin{itemize}
\item [$\bold H_4$)]  $\int^\infty_{-\infty} \Psi(t) \cdot \partial _kF(q(t),
k_\infty)dt > 0$.

\end{itemize}

From Silnikov \cite{silnikov68}, (\ref{E3.3}) and $H_4$) imply that there 
exists $\epsilon>0$ 
so that  for $k_\infty -\epsilon< k< k_\infty$, system (\ref{E1.1}) has a simple
periodic solution $p(t,k)$ which is orbitally near $q(t)$ and is asymptotically
stable.  A more transparent relation indicating that the periodic solutions can
only be found for $k<k_\infty$ with $k-k_\infty = O(e^{-T\lambda _+})$ is given
in \cite{lin90},  where $T$ is the period of $p(t,k)$.  There is a one-to-one
correspondence between $k$ and $T$.  Moreover,
there exists $C>1$, independent of $k$, such that
$$
C^{-1} e^{-T\lambda _+} \leq {\frac {dk} {dT}} \leq C e^{-T\lambda _+}.
$$
The proof of that can be obtained by the same method used in \cite{lin90}.

Consider eigenvalues for the linear variational equation around the equilibrium
$E(k_\infty)$.  It can be verified that each eigenfunction must be in
one ${\cal X}_n,\, n \geq 0$ with an eigenvalue $\lambda_n$ satisfying
$$
\det \left ( \begin{array}{cc} \lambda + a + n^2\pi^2d_1 & b\\
c & \lambda + d+ n^2\pi^2d_2\end{array}\right ) = 0.
$$
The spaces ${\cal X}_n$ are defined in  \S2.
For each ${\cal X}_n$, denote the eigenvalues corresponding to the $n$-th 
Fourier mode by  $(\lambda_{n1}, \lambda _{n2}$), with $\mbox {Re} \lambda _{n1} \geq \mbox { Re } \lambda _{n2}$.  Based on $a+d>0$, we have 
$\mbox {Re } \lambda _{n2} < 0$.  An $n$-th mode
is unstable if and only if $\mbox { Re } \lambda _{n1} > 0$.  The critical
case $\lambda _{n1} = 0$ occurs if
$$
(a+n^2\pi^2d_1)(d+n^2\pi^2d_2) = bc.
$$

We can show that when decreasing $(d_1,d_2)$, the first mode loses stability 
before  the other Fourier modes.
(Theorem 3.1).  Thus, we are interested in parameter values where $\lambda
_{11} = 0$.  Define
$$
\Gamma = \{(d_1,d_2): (a+\pi^2d_1)(d+\pi^2d_2) = bc\}
$$

\begin{thm}\label{T3.1}  The first quadrant, ${\Bbb R}^2_+$, is divided by
$\Gamma$ into two regions:  ${\cal G}_+$ and ${\cal G}_-$ where
$(a+\pi^2d_1)(d+\pi^2d_2)-bc < 0$ and $> 0 $ respectively.

\begin{itemize}
\item [(i)]  $\lambda _{11} > 0$ in ${\cal G}_+$. If $d_1$ and $d_2$ are
sufficiently small, then $(d_1,\,d_2)\in \cal{G}_+$. 

\item [(ii)]  Re $\lambda_{11} < 0$ in ${\cal G}_-$.  The region ${\cal
G}_-$ is unbounded.

\item [(iii)]  $\lambda _{11} = 0$ on $\Gamma$.

\item [(iv)]  $\lambda _{01} = \lambda _+ > 0$ in ${\Bbb R}^2_+$.  If
$(d_1,d_2) \in {\cal G}_-\cup\Gamma$, then Re$\lambda_{nj} < 0$ for
$(n,j)\neq (0,1)$ or $(1,1)$.

\item [(v)]  $\nabla \lambda _{11} = (\partial_{d_1} \lambda _{11},
\partial_{d_2} \lambda _{11}) \neq 0$ for $(d_1,d_2)\in \Gamma$. 
In  particular, $\partial_{d_1}\lambda_{11} < 0$ if $d+\pi^2d_2>0$
and $\partial_{d_2}\lambda_{11} < 0$ if $a+\pi^2 d_1 > 0$.

\end{itemize}
\end{thm}
\vskip1.0in
\centerline{\bf Fig. 3.1}

\thmref{T3.1} will be proved in $\S 6$.
Fig. 3.1 depicts $\Gamma$, ${\cal G}_+$ and ${\cal G}_-$ for all possible cases
except for a possible permutation of $d_1$ and $d_2$. When $bc>0$, we may have
$a>0,\,d>0$ which is Case 2 in Figure 3.1. We may also
have $a>0,\,d\leq 0$ which is Case 1 in Figure 3.1.  It is impossible to have
$a<0,\,d<0$ since $a+d>0$. The other possible
case is $a\leq 0,\,d>0$ which is obtained from Case 1 by symmetry.  When
$bc=0$, since $ad-bc<0$, we have $ad<0$.  Thus we have either $a<0,\,d>0$
which is Case 3 in Figure 3.1, or $a>0,\,d<0$ by symmetry.  When $bc<0$,
again $ad-bc<0$ implies $ad<0$.  Case $a<0,\,d>0$ is in Case 4, the other
case $a>0,\,d<0$ is obtained by symmetry.  Observe in Case 4, when
increasing $d_2$, we can move from ${\cal G}_-$ to ${\cal G}_+$.  It is
interesting to note that the equilibrium may become more unstable by increasing
one of the diffusion coefficient.

The following theorem was stated in \cite{lin92}.

\begin{thm}\label{T3.2}  For each positive $(d_1,d_2) \in {\cal G}_-$,
there exists a smooth function \newline $\epsilon ^*(d_1,d_2)  > 0$ 
such that the SH periodic solution $p(t,k)$ is asymptotically
stable in $D_A(\theta +1)$ if $k_\infty - \epsilon ^* < k < k_\infty$.
\end{thm}

\thmref{T3.2} can be proved by using notions of exponential dichotomies and
roughness of exponential dichotomies in $D_A(\theta +1)$.  
The proof is similar
to the proof of Theorem 4.5 in \cite{lin94}.  Since those methods are quite
different from the ones used in this paper, we will not give details here.

The result in \thmref{T3.2} is not very precise  since $\epsilon^*(d_1,d_2) 
\to 0$ as $(d_1,d_2)\to \Gamma$.  For a given $k$ (or period $T$), the loss of
stability for $p(t,k)$ does not happen exactly at $\Gamma$.

To describe what happens near $\Gamma$, two new notions are introduced: 1)
the stability of the equilibrium for the flow on $W^c_{loc}$; 2) the 
twistedness of the homoclinic
orbit  when following $q(t)$ from $t = -\infty$ to $t =
+\infty$.

When $(d_1,d_2) \in \Gamma$, $\lambda _{11} = 0$, the equilibrium $E(k_\infty)$
has a one-dimensional center manifold $W^c_{loc}$ that is tangent to the
one-dimensional eigenspace  corresponding to $\lambda _{11} = 0$.  The flow on
$W^c_{loc}$ is described in \thmref{T2.7}.  When $\lambda_{11} = 0$, it has
the form
$$
\begin {array}{l}
z^\prime = -\hat cz^3 + \mbox {h.o.t.}\\
x = 0, y = 0.\end{array}
$$

\begin{itemize}
\item [$\bold H_5$)]  When $\lambda_{11} = 0$, the equilibrium $E$ is  stable
on $W^c_{loc} (E)$ in the sense that $\hat c > 0$.
\end{itemize}

Numerical computation in $\S 7$ shows
that in Freedman and Wolkowicz's example the condition $\hat c > 0$ is valid
for all $(d_1,d_2) \in \Gamma$ in the range specified by $0<\pi^2 d_1 < 3$.

Twistedness of the homoclinic solution $q(t)$ has been described in $\S 1$.
Because its importance, we will give a simple and equivalent definition.  Let
$k = k_\infty$ and $(d_1,d_2) \in \Gamma$.  Linearize (\ref{E1.2}) around
$q(t)$ we have
\begin{equation}\label{E3.4}
U^\prime (t) = DU_{\xi\xi}(t) + \partial _UF(q(t),k_\infty)U(t).
\end{equation}
The subspace of the first Fourier mode ${\cal X}_1$ is invariant under 
(\ref{E3.4}).
Since ${\cal X}_1$ is two dimensional, (\ref{E3.4}) on ${\cal X}_1$ reduces to 
an ODE on $(u_1,v_1)$.
\begin{equation}\label{E3.5}
\frac{d}{dt}\left( \begin{array}{c}
 u_1\\ v_1 \end{array}\right) = {\cal A}(t) \left( \begin{array}{c} u_1\\
v_1\end{array}\right), 
\end{equation}
where ${\cal A}(t) \to {\cal A}(\infty) = \left ( \begin{array}{cc}
{-\pi^2d_1-a} & {-b}\\ {-c} &{ -\pi^2d_2-d}\end{array}\right )$ as $t \to \pm
\infty$.  According to a theorem in \cite{hartman64}, each solution of 
(\ref{E3.5}) approaches a solution of the linear autonomous equation,
$$
\frac{d}{dt}\left (\begin{array}{c}
u_1\\ v_1\end{array}\right ) = {\cal A}(\infty) \left (\begin{array}{c} u_1\\
v_1\end{array}\right ),
$$
with an exponentially small error.  Therefore, there is a unique solution
$(u_1(t),v_1(t))$ to (\ref{E3.5}), up to multiplying by scalar constants, that
approaches an eigenvector of the zero eigenvalue of ${\cal A}(\infty)$ as 
$t \to -\infty$. Let
$\tilde U(t) = (u_1(t)\,\cos\, \pi x, v_1(t)\,\cos\,\pi x)$ be the unique
solution of (\ref{E3.4}) that is in ${\cal X}_1$ and approaches an eigenvector 
$\phi_c$, corresponding to $\lambda_{11}=0$ as $t\to
-\infty$.  
By the same argument, when $t\to +\infty,
\tilde U(t)$ approaches another eigenvector associated to $\lambda _{11} = 0$,
denoted by $c^*\phi _c$, where $c^*$ is a function of $(d_1,d_2)\in
\Gamma$.

\begin{defn}\label{D3.3}  Let $\lim_{t\to +\infty} \tilde U(t) = c^*\phi
_c$.  The homoclinic solution $q(t)$ is said to be nontwisted if $c^*>0$, 
twisted if $c^*< 0$, or degenerate if $c^* = 0$.
\end{defn}

\noindent{\bf Remark.}  In $D_A(\theta + 1)$, solutions of (\ref{E3.4}) that 
approach
$\phi _c$ as $t \to -\infty$ are not unique.  They have the form $U(t) =
\tilde U(t) + C\dot q(t)$ where $C$ is an arbitrary constant.  Since $\dot q(t)
\to 0$ exponentially as $t \to +\infty$, we have $\lim _{t\to \infty} U(t) =
c^*\phi _c$ for any $C\in {\Bbb R}$.  Therefore the twistedness defined in
Definition \ref{D3.3} is precisely the one given in $\S 1$.
\vskip 0.05in

Let $k = k_\infty$ and $(d_1,d_2) \in \Gamma$.  From \thmref{T3.1},
$\nabla \lambda _{11} \neq 0$.  It is also obvious that $\nabla
\lambda _{11}$ intersects $\Gamma$ transversely at $(d_1,d_2)$.  
We can
make a smooth change of variable ${\cal B} :(d_1,d_2) \to (\ell,m)$ in a
neighborhood of $\Gamma$ so that $m = \lambda _{11}$ and $\ell$ is the arc
length on $\Gamma$ when $m = 0$, after assigning $\ell = 0$
to an arbitrary point on $\Gamma$.  The new coordinates flatten
$\Gamma$, i.e. $\Gamma = \{m=0, \ell \in \tilde I\}$, where $\tilde I \subset
{\Bbb R}$ is an open interval.

For $m \approx 0$ and $\ell \in \tilde I$, we look for simple period $T$ or
symmetric double period $2T$ SN solutions, where $T>\overline t$, 
$\overline t$ being a
large constant.  In the parameter space $(T,\ell,m)$ we want to find regions
where such SN solutions exist.

For $(d_1,d_2) \in \Gamma$, ${\cal B}(d_1,d_2) = (\ell _0,0)$, and 
the twistedness $c^*=c^*(\ell_0)$ is a function of $\ell_0$.

Throughout this paper, assume that the hypotheses $H_1$)--$H_5$) are satisfied.

\begin{thm}\label{T3.4}  For each $\ell_0 \in \tilde I,\
c^*(\ell_0) \neq 0$, there exist a large constant $\overline t > 0$ and an
open set ${\cal O} \subset {\Bbb R}^2$ containing $(\ell _0,0)$, the size of
which depends on $\ell _0$, such that two $C^1$ functions $L(T,\ell,m)$ and
$r(T,\ell,m)$ can be defined for $T > \overline t$ and $(\ell,m) \in \cal O$,
with the following properties:

\begin{itemize}
\item [1)]  $r(\infty, \ell_0,0) = c^*(\ell_0)$, where $r(\infty,\ell_0,0)
= \lim_{t\to +\infty} r(t,\ell _0,0)$;

\item [2)]  $L(T,\ell,m) = e^{mT} + O(e^{-\alpha T}),\,0<m<\alpha$;

\item [3)]  ${\frac {\partial} {\partial m}}\{L(T,\ell,m)r(T,\ell,m)\}
 > 0$ (or $< 0)$ when $c^*(\ell _0) > 0$ (or $<0)$.
\end{itemize}

Moreover, the existence and uniqueness of simple or symmetric double SN periodic
solutions to (2.1) are determined by the following conditions:

\begin{itemize}
\item [i)]  If $0 < c^*(\ell _0) < 1$ or $1<c^*(\ell _0)$, then there is
no simple period T SN solution when $0 \leq L(T,\ell,m)r(T,\ell,m) \leq 1$;
there are precisely two simple period T SN solutions $U_1(t,\xi)$ and
$U_2(t,\xi)$ when $L(T,\ell,m)r(T,\ell,m) > 1$.  The two solutions are related
by $U_2(t,\xi) = U_1(t,1-\xi)$.

\item [ii)]  If $c^*(\ell_0) = 1$, then there exist two simple period
T SN solutions when \newline  $L(T,\ell,m)r(T,\ell,m) > 1$.  There exists
$\delta>0$ such that the number
of solutions are precisely two when $L(T,\ell,m)r(T,\ell,m)
\geq 1 + \delta$  for some $\delta > 0$;  and there is no simple period T SN
solution when $0\leq L(T,\ell,m)r(T,\ell,m) \leq 1 - \delta$.

\item [iii)]  If $-1 < c^*(\ell _0) < 0$ or $c^*(\ell _0) < -1$, then
there is precisely one SN symmetric double  period 2T solution $U(t,\xi)$
when $L(T,\ell,m)r(T,\ell,m) < -1$.
There is no such SN period 2T solution when $-1 \leq L(T,\ell,m)r(T,\ell,m)
\leq 0$.

\item [iv)]  If $c^*(\ell _0) = -1$, then there is at least one symmetric 
double period 2T SN solution when  $L(T,\ell,m)r(T,\ell,m) < -1$.
Such solution is unique when
$L(T,\ell,m)r(T,\ell,m)\\ < -1 - \delta$ for some $\delta > 0$.  There is no
such SN period 2T solution when $-1 + \delta \leq L(T,\ell,m)r(T,\ell,m)\leq
0$.
\end{itemize}
\end{thm}

\noindent{\bf Corollary.} When $m>0$, there is a pair of SN equilibria $E_1,
\,E_2$ bifurcating from $E$. The results above also show the bifurcation
of SN homoclinic or heteroclinic solutions  asymptotic to $E_1$ and/or $E_2$ 
as a special cases when $T=\infty$. If $c^*\neq 0$, the curve $Lr=1$ is 
identical to $\Gamma$. When crossing $\Gamma$ at a point where 
$c^*(\ell_0)>0$, the bifurcation of a pair of homoclinic solutions, each 
asmptotic to $E_1$ or $E_2$ occurs. When crossing $\Gamma$ at a point where 
$c^*(\ell_0)<0$, the bifurcation of a pair of heteroclinic solutions 
connecting $E_1$ and $E_2$ occurs.

\begin{thm}\label{T3.5}  For each $\ell_0 \in \tilde I$ with $c^*(\ell
_0) = 0$ and $\frac {d}{d \ell} c^* (\ell _0) \neq 0$, there
exist constants  $\epsilon > 0$ and $\overline t > 0$ such that functions $\ell
^*(m),|m| < \epsilon$ and $\delta (T) = ce^{-mT},\,T>\overline t$ for some
$c > 0$ can be defined.  If $|\ell - \ell ^*(m)| < \delta (T),\ |m| <
\epsilon$ and $T >
\overline t$, then there is no simple period T or symmetric double period 2T 
SN solution to (2.1), inside a $(\delta(T))^{\scriptstyle 1/2}$ neighborhood 
of the orbit of $q(t)$.
\end{thm}

\thmref{T3.4} provides fairly accurate information about the bifurcation to
simple or symmetric double periodic SN solutions when crossing the curve
$L(T,\ell,m)r(T,\ell,m) = 1$ not near the points $c^*(\ell _0) = \pm 1$ or 
$c^*(\ell_0)=0$.  First, \thmref{T3.4}, 3) assures that $L(T,\ell,m)r(T,\ell,m)$
is monotonic in term of $m$.  From
the asymptotic forms 1) and 2), it is also clear that the sign of $Lr-1$ 
changes  when $m$ is increased form negative to positive, provided that $T$ is
large. . When crossing the curve $Lr = 1$ near $c^*(\ell
_0) = \pm 1$, bifurcation to simple or symmetric double periodic SN solution will occur
but the precise moment is unknown.  Our method does not predict the existence
or uniqueness of such solutions in a narrow strip around $Lr = 1$.  
\thmref{T3.5}, on the other hand, assures that when $c^*(\ell_0)=0$, we can 
pass $m = 0$
through a small
tubular neighborhood of $\ell = \ell ^*(m)$ without creating any simple period
T or symmetric double period 2T SN solution.  The size of the tubular neighborhood
shrinks to zero as $T \to + \infty$.

The regions in the $(d_1,d_2)$ plane mentioned in \thmref{T3.4} and 
\thmref{T3.5} are depicted in Figure 3.2 where we assume that 
$L(T,\ell,m) = e^{mT},\, r(T,\ell,m) = c^*(\ell)$ and $\ell ^*(m) = 0$.  In the 
shaded area, the existence and uniqueness of a simple period T (or symmetric 
double period 2T) SN solution is guaranteed except near $c^*(\ell) =\pm1$. The 
tubular neighborhood near $\ell = 0$ where crossing $m = 0$ without causing
bifurcation to simple or symmetric double period SN solution is also shown. A 
sketch of all kinds of homoclinic, heteroclinic and periodic solutions is 
in Figure 3.3.
\vskip1.0in
\centerline{\bf Fig. 3.2}

\vskip1.0in
\centerline{\bf Fig. 3.3}

%% file: sect4.tex
\section {Some lemmas}

The results in \lemref{L4.1} is our major tool to study a solution $U(t),\
0 \leq t \leq t_0$, that stays in a small neighborhood of a nonhyperbolic
equilibrium.  Following an idea of Silnikov, we show that if $t_0$ can be
arbitrarily large,
$U(t) = (x(t),y(t),z(t)), \,0\leq t\leq t_0$ is determined by, and depends
continuously on  its boundary values: $y(0),z(0)$ and
$x(t_0)$.  Using exponential dichotomies we can easily show $x(t) =
O(e^{-\alpha (t_0-t)})$ and $y(t) = O(e^{-\alpha t})$ for some $\alpha > 0$.
However, in the center direction, the flow is not exponentially decaying
either moving forward or backward. Following the approach of \cite{chowlin90},
we will compare the $Z$ coordinates of $U(t)$ with a (nonunique) solution
$U_0(t)$ on $W^c_{loc} (E)$.  Let $P_x,\,P_y,\,P_z$ be the spectral projections
from $D_A(\theta+1)$ onto $X,\,Y,\,Z$. In the flat coordinates, we show that
$\|P_z(U(t) - U_0(t))\|$ is small and approaches zero uniformly for
$t \in [0,t_0]$ as $t_0 \to +\infty$.  If we are interested only in
dynamics in the $Z$ direction, $U(t)$ can be replaced by $U_0(t)$ on
$W^c_{loc}(E)$ with a very small error. 

It is also clear that the smallness of $P_z(U_0(t) - U(t))$ strongly depends on
a good choice of coordinates.  Since $x(t_0)$ and $y(0)$ are not small as
$t_0\to \infty$, a undesired change of variables may destroy the smallness of
$P_z(U_0(t) - U(t))$.

Let $\cal O\subset D_A(\theta + 1)$ be a small neighborhood of an equilibrium
$U = 0$ where the flat coordinates introduced in $\S 2$ are used in $\cal O$.
We now consider the abstract parabolic equation (\ref{E2.1}) written in the flat
coordinates,
\begin{equation}\label{E4.1}
\begin{array}{lll}
x^\prime & =&  A_1x + g_1(x,y,z,\mu),\\[6pt]
y^\prime & =&  A_2y+g_2(x,y,z,\mu),\\[6pt]
z^\prime & =&  A_3z + g_3(x,y,z,\mu).
\end{array}
\end{equation}
Here $A_1 = A|_X,\ A_2= A|_Y$ and $A_3 = A|_Z$. Re$\sigma(A_1) > \lambda _M >
0$, Re$\sigma(A_2) < -\lambda _M < 0$ and Re$\sigma(A_3) = 0$.  The
functions $g_i, i = 1,2,3$, are $C^\nu$, $\nu\geq 2$ in all the variables. Since
the coordinates are flat, it can be verified that $g_1(0,y,z,\mu) = 0$,
$g_2(x,0,z,\mu) = 0$ and $g_3(0,y,z,\mu) = g_3(x,0,z,\mu) = g_3(0,0,z,\mu)$.
Moreover, $D_Ug_i(0,0,0,\mu_0) = 0,\ i=1,2,3$. 
The equation  for the flow on the center manifold is
\begin{equation}\label{E4.1a}
z^\prime=A_3 z+g_3(0,0,z,\mu).
\end{equation}
Let $\Phi(t,z_0,\mu)$ be the solution map for (\ref{E4.1a}), with
$\Phi(0,z_0,\mu)=z_0$.
We have the following

\begin{lem}\label{L4.1}  For any $\alpha_0,\beta>0$ with $0<\beta<\alpha_0
< \lambda_M$, there exist
 positive constants $\epsilon_M,\,\delta_M,\, \mu_M$ and
$t_m$ with the following properties.  The constant $\epsilon_M$ is small
enough so that
$\{U=(x,y,z)\mid \|x\|_X \leq\epsilon_M,\ \|y\|_Y \leq\epsilon_M,\
\|z\|_Z \leq\epsilon_M\} \subset \cal O$.
If $|\mu| < \mu_M, t_0 \geq t_m$ and $z_0 \in Z$, satisfying
$$
\|\Phi(t,z_0,\mu)\| _Z \leq\epsilon_M \quad \mbox { for } t \in [0,t_0],
$$
and if $|x_0|+|y_0|<\delta_M$, then there exists a unique solution $U(t) \in
{\cal O},\ t\in [0,t_0]$, to equation (\ref{E4.1}), satisfying the boundary conditions
$$
x(t_0) = x_0,\ y(0) = y_0\mbox { and } z(0) = z_0.
$$

The solution can be written in the form
$$
U(t) = (x^{\$}(t), \ y^{\$}(t),\ \Phi(t) + z^{\$}(t)),\ 0 \leq t \leq t_0,
$$
where $\Phi(t) = \Phi(t,z_0,\mu),\ z^{\$}(0) = 0$. $w^{\$}(t) =
w^{\$}(t;t_0,x_0,y_0,z_0,\mu),\ w = x,y$ or $z$, are $C^{\nu-1}$ functions in 
all
the variables if  $g_i,\ i = 1,2,3$ is $C^\nu$.  Moreover, let $r$ be a 
multi-index with
$0 \leq |r| \leq \nu -1$.  Suppose $\alpha_1$ satisfies $0<\beta <\alpha_1<
\alpha_0-|r|\beta$.  Then
$$
\begin{array}{l}
|D^r x^{\$}(t)|_X \leq Ce^{\alpha _1(t-t_0)},\\
|D^r y^{\$}(t)|_Y \leq Ce^{-\alpha_1t},\\
|D^r z^{\$}(t) \|_Z \leq Ce^{-\alpha _1t_0+\beta t},\ \ 0 \leq t \leq
t_0.\end{array}
$$
\end{lem}

The proof for \lemref{L4.1} in the ODE case can be found in 
\cite{chowlin90,deng90,lin92a}.  The proof for systems of abstract parabolic
equations is similar and will not be rendered here.

Since the small eigenvalue $\lambda_{11}=m$ and since the flow on the center
manifold is odd, we can rewrite (\ref{E4.1a}) as the following, 
\begin{equation}\label{E4.2}
z^\prime = mz - \hat cz^3 + z^5h_1(z,\mu).
\end{equation}
Here $\hat c=\hat c(\mu) > 0$ due to $H_5$), and  $|\mu| \leq \mu_M$.  The
function
$h_1$ is $C^\nu$ for all $\nu>0$ and $h_1(-z,\mu) = h_1(z,\mu)$.  
Equation (\ref{E4.2}) has three
equilibria $z = 0$ and $z = \pm z_E$, where $z _E  \approx \sqrt {m/\hat c}$
provided that $m > 0$ and $m$ is small.  
 
In \lemref{L4.2} and \lemref{L4.3} we present some estimates on  the function
\newline
$\Phi(t,z_0,\mu)/z_0$ which measures the degree of expansion or
contraction on the center manifold. The importance of 
these estimates will be clear in
the next two sections where bifurcation functions and their approximations are
introduced. The proofs are technical and can be skipped on the first reading.
In fact, the results in \lemref{L4.2} and \lemref{L4.3}  are easy to
verify for the truncated equation
$$ z^\prime=mz-\hat c z^3.
$$
All we try to show in these lemmas is that the perturbation term 
$z^5 h_1(z,\mu)$ does not change the solution significantly.

Let $\epsilon>0$ be a small constant. By plotting the phase diagram of 
(\ref{E4.2}) on $(-\epsilon,\epsilon)$,
see Figure 4.1, it can be verified that
$|\Phi(t,z_0,\mu)| < \epsilon$ provided $|z_0| < \epsilon $, and $m$ and $\mu_M$
are small.
In \lemref{L4.2}, we show $\Phi(t,z_0,\mu)/z_0$ is monotonic with respect to 
$z_0$ in $(0,\epsilon)$ if
$t>0$ is fixed. We also give formulas that will provide some lower bounds
on $\left |\frac{\partial}{  \partial z_0} \left ( \frac{\Phi
(t,z_0,\mu)}{ z_0}\right )\right |$ in the future.
\vskip 0.1in
\bg{center} {\bf Fig. 4.1}
\end{center}
 
\begin{lem}\label{L4.2}               
There exists $\epsilon > 0$ such that
$$
\mbox { sign } \left \{ \frac{\partial}{ \partial z_0} \left ( \frac{\Phi
(t,z_0,\mu)}{ z_0}\right ) \right \} = - \mbox {sign}  z_0
$$
if $0 < |z_0| < \epsilon$.  Moreover, we can show  the following.

\begin {itemize}
\item [(i)]  If $m \leq 0$, then
$$
\frac{\partial }{ \partial z_0} \left [ \frac{\Phi }{ z_0}\right ] =
\frac{C_1(z^2_0-\Phi^2)}{ m-\hat cz_0^2 + z^4_0 h_1(z_0,\mu)}\cdot \frac{\Phi}{
z_0^2},
$$
where $C_1$ is a function of $z_0$.  $C_1 \approx \hat c$ if $m$ and $\epsilon$
are small.
\item [(ii)]  If $m \geq 0$ and $z^2_0 \neq z^2 _E$, then we have
$$\frac{\partial }{\partial z_0} \left [ \frac{\Phi}{ z_0}\right ] =
\frac{C_2(z^2_0 - \Phi ^2)}{ z^2_E - z^2_0}\cdot \frac{\Phi }{ z^2_0},
$$
where $C_2$ is a function of $z_0$.  $C_2 \approx 1$ if $m$ and $\epsilon$ are
small.
\item [(iii)]  If $m > 0$ and $z^2_0 = z^2_E$, then
$$
\frac{\partial }{ \partial z_0} \left [ \frac{\Phi }{ z_0}\right ] =
(e^{-m^\prime t} - 1)/z_0,
$$
where $-m^\prime = \frac{\partial }{ \partial z} \left [ mz-\hat  cz^3 +
z^5h_1(z,\mu)\right ] \mid _{z=z_E} \approx -2m$ if $m$ and $\epsilon$
are small.
\end{itemize}
\end{lem}

\bg{pf}  Since $\Phi (t,z_0,\mu)$ is an odd function of $z_0$,
it suffices to consider $z_0 > 0$.  Let $z = \Phi (t,z_0,\mu)$.  It is well 
known that $U(t)=\frac{\partial \Phi }{ \partial z_0}$ satisfies the linear variational
equation for (\ref{E4.2}) with $U(0)=1$, so does 
$\frac{\partial\Phi(t,z_0,\mu)}{
\partial t}/\frac{\partial\Phi(0,z_0,\mu)}{ \partial t}$. Therefore, they must
be identical. Using (\ref{E4.2}) to replace $\frac{\partial\Phi(t,z_0,\mu)}{
\partial t}$, we have
\begin{align}\label{E4.3}
\frac{\partial \Phi}{ \partial z_0}  &= \frac{mz - \hat cz^3 +
z^5h_1(z,\mu)}{ mz_0 -\hat cz^3_0 + z^5_0h_1(z_0,\mu)}.\\
\frac{\partial }{ \partial z_0} \left ( \frac{\Phi }{
z_0}\right)
& = \left \{ \frac{mz -\hat cz^3 + z^5h_1(z,\mu)}{ mz_0 -\hat cz^3_0 + z^5_0
h_1(z_0,\mu)} \frac{z_0}{ z} - 1\right \} \frac{z}{ z^2_0} \label{E4.4} \\
& = \frac{\hat c(z^2_0 - z^2) + z^4h_1(z,\mu) -
z_0^4h_1(z_0,\mu)}{ m-\hat cz^2_0+z^4_0h_1(z_0,\mu)} \cdot \frac{z }{ z^2_0}
\notag
\end{align}
Since $h_1(z,\mu)$ is an even function of $z$, we have
$$
z^4h_1(z,\mu) - z^4_0h_1(z_0,\mu) = C_3(z^2_0 - z^2),
$$
where $C_3$ is a function of $z_0$ and is small if both $z$ and $z_0$ are small.This proves (i).

For any fixed $t > 0,\ m > 0$, let $z_0\to z_E$, then $z=\Phi\to z_E$. From 
(\ref{E4.3}), and the fact $\frac{\partial \Phi }{ \partial
z_0} \to e^{-m^\prime t}$ as $z_0 \to z_E$, $t$ fixed, we have
$$
\lim_{z_0\to z_E} \frac{mz-\hat cz^3 + z^5h_1(z,\mu)}{ mz_0 - \hat cz^3_0 +
z^5_0h_1(z_0,\mu)} = e^{-m^\prime t},
$$
where $-m^\prime = \frac{\partial }{ \partial z} [mz -\hat cz^3 +
z^5h_1(z,\mu)]|_{z=z_E} \approx - 2m$, since $z_E \approx \sqrt {m/\hat c}$.
Therefore, (iii) follows from the first line of (\ref{E4.4}).

When $m > 0$, since $z_E$ is a nonzero equilibrium, $m-\hat cz_E^2 + 
z^4h_1(z_E,\mu) = 0$.  Therefore
$$
\begin{array}{ll}
&m-\hat cz^2_0 + z^4_0h_1(z_0,\mu)\\
=& m -\hat cz^2_0 + z^4_0h_1(z_0,\mu) - [m -\hat cz^2_E + z^4_Eh_1(z_E,\mu)]\\
=& C_4(z^2_E - z_0^2),\end{array}
$$
where $C_4$ is a function of $z_0$ and is close to $\hat c$.  From this,
(ii) follows from the second line of (\ref{E4.4}).

When $m \leq 0$, zero is an attractor on $W^c_{loc}$. If $ z_0 > 0$, then 
$z^2_0 > \Phi^2$.  Thus $\frac{\partial }{
\partial z_0} [\frac{\Phi }{ z_0}]<0$, based on (i).  When $m>0,\ z_0=z_E,\
\frac{\partial }{ \partial z_0} [\frac{\Phi }{ z_0}] < 0$, based on (iii).  
When $m > 0,\ z_0 \neq z_E,\, z_0 > 0$, 
$z_E$ attracts $z_0$. From the phase diagram,
see Figure 4.1, it is clear that $z^2_0 -\Phi^2$ and $z^2_E - z^2_0$ always 
have different signs.  
Thus $\frac{\partial }{ \partial z_0} [\frac{\Phi }{ z_0}] < 0$, based on (ii).
\end{pf}

In the next lemma, we derive some estimates on the rate of contraction or 
repelling for the equilibria zero and/or $\pm z_E$ on the center manifold.
These estimates are to be used in conjunction with \lemref{L4.2}.
 
\begin{lem}\label{L4.3}
There exist $\overline z>0$ and $\overline m>0$ with the following
properties.  \newline
(a) Let $|z_0|<\overline z,\,0<m<\overline m$ and $\Phi=\Phi(t_0,z_0,\mu)$.
For any $\tau_0>0$, if $mt_o>\tau_0>0$ then there
exists $\eta=\eta(\tau_0)>0$ such that either $$z_o^2\leq(1-\eta)\Phi^2
\mbox{ or } |\Phi^2-z_E^2|\leq(1-\eta)|z_0^2-z_E^2|.$$
(b) If $-\overline m\leq m<0$ and $-mt_o>\tau_0$, then
$$\Phi^2\leq(1-\eta)z_0^2$$
\end{lem}

\bg{pf} Let $w=z^2$. Define $h(w,\mu)=2mw-2\hat cw^2+
2w^3h_1(\sqrt{w},\mu)$. We have $w'=h(w,\mu)$. Let the solution map be
$w(t,w_0)$.  Since $h_1$ is a $C^\infty$,  even function of $z$,
it can be shown that $h$ is $C^\infty$.  Let $\hat w
=\frac{m}{ 3\hat c}$ and $w_E=z_E^2=\frac{m}{\hat c}+O(m^2)$. It is easy to see
that $\frac{\partial^2h}{\partial w^2}<0$ if $\overline z$ is small and 
$w<\overline z^2$.  Therefore, using 
Taylor's formula with remainder, we have
$$\frac{\partial}{\partial w}\left(\frac{h(w,\mu)}{ w}\right)=
\frac{\frac{\partial}{\partial w}h\cdot w- h}{ w^2}<0.$$
Here we have used the fact that $h(0,\mu)=0$.  Similarly, since $h(w_E,\mu)
=0$, $$\frac{\partial}{\partial w}\left(\frac{h(w,\mu)}{ w-w_E}\right)<0.$$

Consider case (a), $m>0$ first.\newline
\noindent (i) If $w_0>w_E$, then $w(t)=w(t,w_0)>w_E$ for all $t>0$. 
Since $h(w_E, \mu)=0$ and $\frac{\partial}{\partial w}(\frac{h}{w})<0$,
$$\frac{h(w,\mu)}{ w-w_E}\leq\frac{\partial}{\partial w}h(w_E,\mu)=-2m+O(m^2)
<-m,$$ if $0<m<\overline m$.  Let $e^{-\tau_0}=1-\eta$. From $(w-w_E)'\leq
-m(w-w_E)$, we have
\begin{eqnarray*}
w(t)-w_E&\leq& e^{-mt_0}(w_0-w_E)\\
&\leq& e^{-\tau_0}(w_0-w_E)\\
&\leq& (1-\eta)(w_0-w_E).
\end{eqnarray*}

\noindent(ii) Observe that $\hat w<w_E$.  If $0<w(\frac{t_0}{ 2},w_0)\leq \hat w$,
then $0<w(t)\leq \hat w$, for $0\leq t\leq \frac{t_0}{2}$. Since $\frac{h}{w}$
is monotonic,
\begin{eqnarray*} \frac{h(w,\mu)}{ w}&\geq& \frac{h(\hat w,\mu)}{\hat w}\\
&=&2m-2\hat c\hat w+O(\hat w^2)\\
 &=&2m-\frac{2m}{ 3}+O(m^2)\\ &>&m,
 \end{eqnarray*}
if $m$ is small.  Let $1-\eta=e^{-\tau_0/2}$.  From $w'\geq mw$, we have
$$w(t_0)>w(\frac{t_0}{ 2})\geq e^{mt_o/2}w_0.$$
Therefore $w_0\leq(1-\eta)w(t_0)$.

\noindent(iii) If $\hat w\leq w(\frac{t_0}{ 2})<w_E$, then for
$\frac{t_0}{ 2}\leq t \leq t_0$, $\hat w\leq w(t)<w_E$. Observe that
$\frac{w_E}{\hat w}= 3+O(m)$. Using the monotonicity of $h/w$,
\begin{eqnarray*} \frac{h(w,\mu)}{ w-w_E}&\leq&\frac{h(\hat w,\mu)}{
\hat w-w_E}\\
&=&\left(\frac{\hat w}{\hat w-w_E}\right)(2m-2\hat c\hat w+ O(\hat w^2))\\
&=&\frac{1}{ -2+O(m)}(\frac{4m}{ 3}+O(m^2))\\
&=&-\frac{2m}{ 3}+O(m^2)\\
&<&-\frac{m}{ 2},
\end{eqnarray*}
if $m$ is small. Let $1-\eta=e^{-\tau_0/4}$. From $(w-w_E)'\geq-\frac{m}{ 2}
(w-w_E)$, we have
\begin{eqnarray*}|w(t_0)-w_E|&\leq&|w(\frac{t_0}{ 2})-w_E|
 e^{-{m\over2}{t_0\over2}}\\
&\leq&|w_0-w_E|e^{-\tau_0/4}\\
&\leq&|w_0-w_E|(1-\eta).
\end{eqnarray*}

Case (b), $m<0$ can be proved similar to case (a), (i).
\end{pf}

\begin{lem}\label{L4.4}  Assume that $ad - bc<0,\ d_1>0,d_2>0$ and
$(a+\pi^2d_1)(d+\pi^2d_2) \geq bc$.  Then $f(\xi) = (a+\xi \pi^2d_1)(d+\xi
\pi^2d_2) - bc$ satisfies $f^\prime (\xi) > 0$ for all $\xi
\geq 1$ and $f(\xi) > 0$ for all $\xi >1$.
\end{lem}

\bg{pf}  The assumption implies that $f(1) \geq 0$.  It is easy
to verify that
$$\begin{array}{ll}
{d\over d\xi} f(\xi) &= 2\xi \pi^4d_1d_2 + a\pi^2d_2+d\pi^2d_1\\
& >  {\frac{1}{\xi}} \{ (a+\xi \pi^2d_1)(d+\xi\pi^2d_2)-bc\}\\
&>  {\frac{1}{\xi}} f(\xi).\end{array}
$$
From this the desired result follows.
\end{pf}

The following lemma relates the Hypothesis $H_4$) with the breaking of the
homoclinic orbit $q(t)$.  It is a variation of a well known result on the
homoclinic bifurcation using Melnikov's integral. See \cite{lin90}.

\begin{lem}\label{L4.5}  
Consider the ODE system (\ref{E1.1}).  Let $\Sigma$ be a
cross section intersecting the orbit of $q(t)$ transversely.  Let $q(0)\in
\Sigma$.  Assume $T_{q(t)}W^u(E) \cap T_{q(t)}W^s(E)$ is one dimensional --
spanned by $\dot q(t)$.  
Let $t_1>0$ and ${\stackrel{\rightharpoonup} {v}} \perp \{T_{q(t_1)} W^u(E) +
T_{q(t_1)} W^s(E)\}$.  Then for each $k \approx k_\infty$,
there exist a unique $g(k) \in {\Bbb R}$ and a piecewise smooth solution 
$U(t,k)$ of (\ref{E1.1})
that is $C^1$ in $(-\infty,t_1)\cup (t_1,\infty)$. Moreover, $U(0,k) \in \Sigma 
\cap W^u(E)$ and
$U(t_1^+,k)\in W^s(E)$ with $U(t_1^+,k)-U(t_1^-,k)=
g(k){\stackrel{\rightharpoonup}{v}} $.  Here $U(t_1^-,k)$ and $U(t_1^+,k)$ 
denote the left and right limit at $t_1$.
Finally, if $H_4$) is valid, then ${d\over dk} g(k) \neq 0$.
\end{lem}

%% file: sect5.tex
 \section {Bifurcation equations for simple and symmetric double periodic 
solutions.}

Let $\overline {x} > 0$ be a small constant and $\Sigma = \{x = \overline
{x}\}$ be a cross section that intersects the orbit of $q(t)$ transversely at
$(\overline {x}, 0,0) \in {\cal O}$. Assume that $q(0)\in\Sigma$. Trajectories 
near the homoclinic orbit
must hit $\Sigma \cap {\cal O}$ at least once.  We can make $\Sigma$ smaller
so that trajectories starting from $\Sigma $ must reenter ${\cal O}$
after a fixed time $t_1$. The cross section $\Sigma$ is used to fix the phase.
We are not construcing a Poincar\'e mapping: $\Sigma\to \Sigma$.

First consider a simple periodic solution of
period $T = t_0 + t_1$. Since $t_1$ is fixed, the period $T$ is determined
by $t_0$.  The solution can be divided into an outer solution $U_*(t) =
(x_*(t),y_*(t),z_*(t))$, $0 \leq t \leq t_1$ and an inner solution $U^*(t) =
(x^*(t),\ y^*(t),\ z^*(t))$, $0 \leq t \leq t_0$.  In the sequel, we use 
superscript (subscript) to denote inner (outer) solutions.
Let the outer solution be specified by 
an initial value problem with the initial value $U_*(0) = 
(\overline x, y_1,z_1)\in \Sigma$ and let the solution be denoted by
$U_*(t;\overline {x}, y_1,z_1,\mu)$.  Let the inner solution be specified by
the boundary
value problem as in \lemref{L4.1} with the boundary
conditions $x^*(t_0) = \overline {x},\ y^*(0) = y_0$ and $z^*(0) = z_0$, and
stays in ${\cal O}$ for all $t\in [0,t_0]$.  See Figure 5.1. By \lemref{L4.1}, 
such inner solution is unique and is denoted by
$$
\begin{array}{ll}
x^*(t) & = \  x^{\$}(t;t_0,{\overline x},y_0,z_0,\mu),\\
y^*(t) & = \ y^{\$}(t;t_0,{\overline x},y_0,z_0,\mu),\\
z^*(t) & = \ \Phi(t;z_0,\mu) + z^{\$}(t;t_0,{\overline x},y_0,z_0,\mu).
\end{array}
$$

Define
$$
\begin{array}{ll}
x^*(t_0,y_0,z_0,\mu) & = \ x^*(0),\\[8pt]
y^*(t_0,y_0,z_0,\mu) & = \ y^*(t_0),\\[8pt]
z^*(t_0,y_0,z_0,\mu) & = \ z^*(t_0),\\[8pt]
{\hat x} (y_1,z_1,\mu) & = \ x_*(t_1;{\overline x},y_1,z_1,\mu),\\[8pt]
{\hat y}(y_1,z_1,\mu) & = \ y_*(t_1;{\overline x},y_1,z_1,\mu),\\[8pt]
{\hat z}(y_1,z_1,\mu)  &=   \ z_*(t_1;{\overline x}, y_1,z_1,\mu).
\end{array} $$

\vskip1.0in
\centerline{\bf Fig. 5.1}

The end points of outer and inner solutions must match.  We have the following
equations
\begin{equation}\label{E5.1}
G_1 {\buildrel \rm def \over =}  \hat x(y_1,z_1,\mu)
- x^*(t_0,y_0,z_0,\mu) = 0,
\end{equation}
\begin{equation}\label{E5.2}
y_1 = y^*(t_0,y_0,z_0,\mu),
\end{equation}
\begin{equation}\label{E5.3}
z_1 = z^*(t_0,y_0,z_0,\mu),
\end{equation}
\begin{equation}\label{E5.4}
y_0 = \hat y(y_1,z_1,\mu),
\end{equation}
\begin{equation}\label{E5.5}
z_0 = \hat z(y_1,z_1,\mu).
\end{equation}
See Figure 5.1.

Substituting (\ref{E5.4}) and (\ref{E5.5}) into (\ref{E5.2}), we have
\begin{equation}\label{E5.6}
y_1 = y^* (t_0,\ \hat y(y_1,z_1,\mu),\ \hat z(y_1,z_1,\mu),\mu ).
\end{equation}
Using the smallness of $\partial  y^*/\partial y_0$ and
$\partial y^*/\partial z_0$, see \lemref{L4.1}, we solve $y_1$ from (\ref{E5.6})
by a contraction principle to yield
\begin{equation}\label{E5.7}
y_1 = \tilde y(t_0,z_1,\mu).
\end{equation}

\begin{lem}\label{L5.1}{} 
\begin {itemize}
\item [i)] There exists a constant $\alpha_1>0$ such that $|\tilde y|+\left|{\partial \tilde y\over \partial z_1}\right|= O(e^{-\alpha_1t_0})$.

\item [ii)] $\tilde y(t_0,-z_1,\mu) = R\tilde y(t_0,z_1,\mu)$.

\item [iii)]  When $z_1 =0,\ ({\overline x} , \tilde y(t_0,0,\mu),0) \in
{\cal X}_0$.
\end{itemize}
\end{lem}

\bg{pf} i)  follows from \lemref{L4.1}.  

Since
$(x_*(t),y_*(t),z_*(t))$ satisfies initial values 
$(x_*(0),y_*(0),z_*(0)) = ({\overline x},y_1,z_1)$, then
$(Rx_*(t),\, Ry_*(t),\, Rz_*(t))$ satisfies initial values 
$(R{\overline x}, Ry_1,Rz_1)$. Therefore
$$
w_*(t_1;R\overline x,Ry_1,Rz_1,\mu)=Rw_*(t_1;\overline x,y_1,z_1,\mu),
$$
where $w_*=x_*,\,y_*$ or $z_*$.  
Next, since $(Rx^*(t), Ry^*(t), Rz^*(t))$ satisfies boundary values
$Rx^*(t_0)$ $ = R{\overline x}=\overline x,\, Ry^*(0) = Ry^0, \ Rz^*(0) = Rz^0$,
$$
w^*(t_0,Ry^0,Rz^0,\mu) = Rw^*(t_0,y^0,z^0,\mu),
$$
where $w^* = x^*, y^*$ or $z^*$.  
Based on these facts, using the uniqueness
of the fixed point, (ii) can be verified from (\ref{E5.6}).  

When $z_1=0$, in (\ref{E5.6}), let $y_1\in{\cal X}_0$.  Since ${\cal X}_0$ is
invariant under the flow, then $y_*\in {\cal X}_0$
and $z_*\in {\cal X}_0$, i.e., $z_*=0$. Since we can solve the boundary
value problem,  as described in \lemref{L4.1}, in ${\cal X}_0$, therefore, 
the right hand 
side of (\ref{E5.6}), i.e., $y^*$ is in ${\cal X}_0$. We then can solve 
(\ref{E5.6}) 
by the contraction principle in ${\cal X}_0\cap Y$.  This implies that
the unique solution $\tilde y(t_0,0,\mu)\in {\cal X}_0$.  iii) then
follows from \thmref{T2.7}.
\end{pf}

We now substitute (\ref{E5.7}) into (\ref{E5.1}).  Recall that 
$\mu=(\ell,m,k).\  G_1$ is now a function of $(t_0,\ell,m,k,z_1)$.
\begin{equation}\label{E5.8}
G_1(t_0,\ell,m,k,z_1) = \hat x(\tilde y(t_0,z_1,\mu),z_1,\mu) -
x^*(t_0,\hat y,\hat z,\mu)
\end{equation}
where the arguments of $\hat y$ and $\hat z$ are
$(\tilde y(t_0,z_1,\mu),z_1,\mu)$.

\begin{lem}\label{L5.2}  i)  $G_1(t_0,\ell,m,k,-z_1) =
G_1(t_0,\ell,m,k,z_1)$.

ii)  ${\partial \over \partial k} G_1\neq 0.$
\end{lem}

\bg{pf}  i)  The functions $\tilde y,\hat y,\hat z,\hat x$ and $x^*$
are all invariant under the reflection $R$.  Therefore $G_1(t_0,\ell,m,k,Rz_1)
= RG_1(t_0,\ell,m,k,z_1)$.  Assertion i) then follows from the facts
$Rz_1 = -z_1$ and $RG_1 = G_1$.

ii)  Set $t_0 = + \infty, z_1 = 0$ and $\mu = \mu _0$ where $\mu _0= (\ell
_0,m_0,k_\infty)$ with $m_0=0$ (or $(\ell_0,m_0)\in\Gamma$).  We then have 
$x^* = 0$ from  \lemref{L4.1} and $\tilde y = 0$ from \lemref{L5.1}.  
We now show that the function $G_1(\infty,\ell _0,0,k,0)$ is 
the Melnikov function as in \lemref{L4.5}.

Since ${\cal X}_0$ is invariant under systems (2.1), and
$({\overline x}, 0,0) \in {\cal X}_0$, we have
$x_*(t_1,{\overline x},0,0,\mu)$ $ \in{\cal X}_0$. In ${\cal X}_0$, the
equilibrium $E$ is hyperbolic with $W^s_{loc}(E) = \{x = 0\},\, W^u_{loc}(E)
= \{y=0\}$. Thus $(\bar x,0,0)\in W^u_{loc}(E)$. Consequently, $U(t_1)$, with
the initila condition $(\bar x,0,0)$ is in $W^u(E)\cap {\cal X}_0$. Observe
that $(1,0,0)$ is
a vector orthogonal to $\dot q(t_1)$ where $t_1$ is a 
large constant such that $q(t_1)$ has reentered ${\cal O}$.  Thus,
$G_1(\infty,\ell_0,0,k,0)$ is the function $g(k)$ in \lemref{L4.5}.  From 
\lemref{L4.5} and Hypothesis
$H_4$), we have ${\partial \over \partial k}G_1(\infty,\ell _0,0,k_\infty,0)
\neq 0$.  Observe that $G_1$ is a $C^1$ function in a neighborhood of
$(\infty,\ell _0,0,k_\infty,0)$. Thus, ${\partial \over \partial k}G_1 \neq
0$ for $(t_0,\ell,m,k,z_1)$ near $(\infty,\ell _0,0,k_\infty,0)$. 
\end{pf} 

Since  $G_1(\infty,\ell_0,0,k_\infty,0) = 0$, reflecting the existence of the
homoclinic solution $q(t)$ at $k = k_\infty$, we can use \lemref{L5.2}, ii) to
solve $k = k^*(t_0,\ell,m,z_1)$ from (\ref{E5.8}), if $t_0  \approx \infty, \ell
\approx \ell_0$, $m \approx 0$ and $z_1 \approx 0$.  From \lemref{L5.2}, i),
\begin{equation}\label{E5.9}
k^*(t_0,\ell,m,-z_1) = k^*(t_0,\ell,m,z_1).
\end{equation}

We now substitute $k = k^*(t_0,\ell,m,z_1)$ into (\ref{E5.3}), to obtain a 
bifurcation function
\begin{equation}\label{E5.10}
G_2(t_0,\ell,m,z_1) {\buildrel \rm def \over = } z^*(t_0,{\hat y}(\tilde
y(t_0,z_1,\mu),z_1,\mu),{\hat z}(\tilde y(t_0,z_1,\mu),z_1,\mu),\mu),
\end{equation}
where $\mu = (\ell,m,k^*(t_0,\ell,m,z_1))$.  A solution  of the equation
\begin{equation}\label{E5.11}
z_1 = G_2(t_0,\ell,m,z_1)
\end{equation}
corresponds to a simple period $T = t_0 + t_1$ solution to (2.1).

\begin{lem}\label{L5.3}  $G_2(t_0,\ell,m,-z_1) = -G_2(t_0,\ell,m,z_1)$.  In
particular $G_2(t_0,\ell,m,0) = 0.$  The solution corresponds to $z_1 = 0,\ k =
k^*(t_0,\ell,m,0)$ is a period $T = t_0 + t_1$ SH solution.
\end{lem}

\bg{pf}  Since the functions $\tilde y,\ {\hat y},\
{\hat z}$ and $z^*$, in the definition of $G_2$ are all
invariant under the symmetry
$R$, so is $G_2$.  since $G_2 \in Z$, we have $RG_2 = -G_2$.  This proves that
$G_2$ is an odd function of $z_1$.

As in the proof of \lemref{L5.2}, $\tilde y(t_0,0,\mu) \in {\cal X}_0$.  The 
outer
solution with initial condition $({\overline x},\tilde y,0) \in {\cal X}_0$
must be in ${\cal X}_0$.  Therefore the periodic solution corresponding
to $z_1 = 0$ is in ${\cal X}_0$. 
\end{pf}

Next, we consider a symmetric double periodic solution $U(t)$ of period 2T.
The bifurcation equation for the existence of such solution can be derived
much the same way as for the simple period T solution.  Therefore we only
discuss it  briefly.
From our definition, $U(t+T) = RU(t), t\in {\Bbb R}$.  Assuming that
$U(0) \in \Sigma = \{x = {\overline x}\}$, we  define
$$
\begin{array}{llll}
U_*(t)& =& U(t),\quad & 0 \leq t \leq t_1,\\
U^*(t)& =& U(t+t_1),\quad & 0 \leq t \leq t_0,\\
T& =& t_0+ t_1.& \end{array}
$$
The matching conditions on the outer and inner solutions are
$$\begin{array}{lcl}
U^*(0)& =& U_*(t_1),\\
U^*(t_0)& =& RU_*(0).\end{array}
$$

As before, let the outer solution $U_*(t)$ be determined by the initial value
$U_*(0) = ({\overline x}, y_1,z_1)$ and the inner solution $U^*(t)$ be
determined by the boundary condition $(x^*(t_0),y^*(0),z^*(0)) = ({\overline
x},y_0,z_0)$.  Then we still have (\ref{E5.1}), (\ref{E5.4}) and (\ref{E5.5}), 
but (\ref{E5.2}) and (\ref{E5.3}) change to
$$
y_1 = Ry^*(t_0,y_0,z_0,\mu),\eqno(\ref{E5.2})'
$$
$$
z_1 = Rz^*(t_0,y_0,z_0,\mu).\eqno (\ref{E5.3})'
$$
Notice that the functions $Ry^*$ and $Rz^*$ have similar smallness
properties like $y^*$ and $z^*$.  We will use the same notations as we
worked on simple periodic solutions if no confusion should
occur.  As before, we can solve $y_1$ to get (\ref{E5.7}), and \lemref{L5.1} is 
still 
valid. Here and afterward we use the same notations for functions $\tilde
y,G_1,k^*,G_2$ when deriving bifurcation equations for both simple and symmetric
double periodic solutions.  Define $G_1$ as in (\ref{E5.8}).  Again we have 
\lemref{L5.2}.
Solving $k = k^*(t_0,\ell,m,z_1)$ as before, we still have (\ref{E5.9}).  
Substituting $k = k^*$ into (\ref{E5.3})${}^\prime$, and defining the function
$G_2(t_0,\ell,m,z_1)$ as in (\ref{E5.10}), we have found that
the solutions of the equation
$$
-z_1 = G_2(t_0,\ell,m,z_1)\eqno (\ref{E5.11})'
$$
correspond to symmetric double period 2T solutions to (2.1).

Analogous to \lemref{L5.3}, we have
$$
G_2(t_0,\ell,m,-z_1) = -G_2(t_0,\ell,m,z_1).
$$
However, when $z_1 = 0$, we really have obtained a SH simple period T
solution tracing its orbit twice, since $U(t+T) = RU(t) = U(t)$ in this case.

%% file: sect6.tex
\section { Proof of the main results.}

\bg{pf*} {\bf Proof of \thmref{T3.1}}  Since $\lambda _{11} \lambda _{12} =
(a+\pi ^2d_1)(d + \pi ^2d_2)-bc$, which is negative (positive or zero) in
${\cal G}_+({\cal G}_-\mbox { or } \Gamma)$,  assertions (i), (ii) and
(iii) follow from the fact Re$\lambda_{12} < 0$.

Let $(d_1,d_2)\in \Gamma \cup {\cal G}_-$. From \lemref{L4.4},
$\lambda _{n1} \lambda _{n2} = (a+n^2\pi ^2d_1)(d+n^2\pi^2d_2) - bc$
is an increasing function for $n \geq 1$.  It follows that
$\lambda_{n1}\lambda_{n2} > 0$ for $n \geq 2$. From Re$\lambda_{n2} < 0$
we have  Re$\lambda_{n1} < 0,\ n \geq 2$. This proves (iv) for $n\geq
2$. The proof for the cases $n = 0,1$ are obvious and will be omitted.

Let $\alpha {\buildrel \rm def \over =} a+\pi^2d_1$ and $\beta {\buildrel \rm
def \over =} d+ \pi ^2d_2$.  $\alpha + \beta > 0$ since $a+d >0$.
$$
\lambda _{11}, \lambda _{12}={\frac {1}{2}}
\left \{ - (\alpha + \beta) \pm \sqrt {(\alpha-\beta)^2+4bc}\right \}.
$$
$$
{\partial \over \partial \alpha} \lambda _{11} =
{1\over 2}  \{ - 1 + (\alpha - \beta)[(\alpha -\beta)^2 + 4bc]^
{-{1\over 2}}\ \}.  $$
When $(d_1,d_2)\in \Gamma$, we have $\alpha \beta - bc = 0$.  Therefore
${\partial \over \partial \alpha} \lambda _{11} = {\frac {1}{2}} \left \{ - 1 +
(\alpha - \beta)/(\alpha + \beta)\right \}$ $ =-{\beta /(\alpha + \beta)}$.
Thus, ${\partial \lambda_{11}\over \partial d_1} < 0$ if $\beta > 0$.  Similarly
${\partial \lambda _{11}\over \partial d_2} < 0$ if $\alpha   > 0$.  It is
impossible to have both $\alpha   \leq 0$ and $\beta \leq 0$  since $\alpha +
\beta > 0$.  This proves (v).
\end{pf*}

As mentioned earlier, we use only one set of notations $\tilde y, k^*,
G_1$ and $G_2$ for functions employed when deriving bifurcation
functions for both simple and symmetric double periodic solutions.  This allows us
to treat both problems simultaneously.

Since the proof of the main results are technical, it may be useful to preview 
the main idea used here. Consider finding a simple SN periodic solution.
Recall that the bifurcation equation 
$$z_1=z^*(t_0,\hat y,\hat z,\mu),
$$
where $\hat y$ and $\hat z$ are as in (\ref{E5.10}), has a trivial solution
$z_1=0$, which corresponds to a SH solution. Since we are not interested
in such solution, it is reasonable to look for solutions of the equation 
$z^*/z_1=1$. 

Let $\mu = (\ell,m,k^*(t_0,\ell,m,z_1))$ and let
\begin{equation}\label{E6.1}
z_0  = {\hat z}
 (\tilde y(t_0,z_1,\mu),z_1,\mu),
\end{equation}
Then $z_0 = 0$ if $z_1 = 0$. We look for solutions of
$${z^* \over z_0}\cdot{z_0 \over z_1}=1.
$$
We can show, in the limiting case, $z^*/z_0 \approx \Phi/z_0$, the latter is the
rate of expansion on the local center manifold. Also $z_0/z_1 \approx 
c^*(\ell)$, the latter is the rate of expansion along the outer solution, and 
its
sign represents the twistedness of the homoclinic orbit.  In particular, based
on \lemref{L4.2}, we can show that
$\frac{z^*}{z_0}$ is monotonic for $z_0>0$ or
$z_0<0$ (\lemref{L6.2}).  The proof of our main results
would be easier if the outer solution were the multiplication by $c^*$ 
and the inner solution were the expansion by the rate $\Phi/z_0$. However, 
such approximations
have some small errors. Care must be exercised to ensure that the error terms do
not disturb the main terms. 

Denote
\begin{equation}\label{E6.2}
r_1(t_0,\ell,m) = \left \{ \lim _{z_1\to 0} {z_0/z_1}\right \}.
\end{equation}
Then
$$
r_1(t_0,\ell,m) = {\partial {\hat z}\over \partial y_1} {\partial
\tilde y\over \partial z_1} (t_0,0,\mu) + {\partial {\hat z}(\tilde
y,0,\mu)\over \partial z_1} + {\partial {\hat z}\over \partial k} \cdot
{\partial k^*(t_0,\ell,m,0)\over \partial z_1}.
$$

Assuming now $t_0 = + \infty$, we have $\tilde y = 0$ and
${\partial \tilde y\over \partial z_1} = 0$ (\lemref{L5.1}).  Also
$k^*(\infty,\ell_0,0,0) = k_\infty$, reflecting the existence
of the homoclinic solution $q(t)$ at $k = k_\infty$. From (\ref{E5.9}), we
can show that  ${\partial \over \partial z_1} k^*(\infty,\ell,m,0) = 0$.
We have shown the first part of the the following

\begin{lem}\label{L6.1}  $r_1(\infty,\ell _0,0) = {\partial
{\hat z}  \over \partial z_1} (0,0,(\ell _0,0,k_\infty))=c^*(\ell _0)$, 
where $c^*(\ell _0)$ is defined in Definition \ref{D3.3}.
\end{lem}

\lemref{L6.1} offers an easy way to compute $r_1(\infty,\ell _0,0)$ since
$c^*(\ell _0)$ can be obtained by computing the ODE system (\ref{E3.5}) in 
${\cal X}_ 1$.  The proof of the second equality of \lemref{L6.1} is deferred 
to $\S 7$.

In the first part of this section we assume that $c^*(\ell _0) \neq 0$ for
an $\ell _0 \in \tilde I$.  Then, based on \lemref{L6.1}, $\frac{\partial z_0}
{\partial z_1}\neq 0$ when $z_1=0$.  We can solve $z_1$ form (\ref{E6.1}) to 
obtain the inverse function
\begin{equation}\label{E6.3}
z_1 = \tilde z_1(t_0,\ell,m,z_0).
\end{equation}
Here $\tilde z_1$ is a smooth function defined for $t_0 \approx + \infty,\ m \approx 0,\ \ell \approx \ell _0$.  
Assume that the domain of $\tilde z_1$ is so small that
$$
\mbox {sign} \{\tilde z_1/z_0 \mid z_0\neq 0\} = \mbox {sign } c^*(\ell _0).
$$

Let
\begin{equation}\label{E6.4}
L_1(t_0,\ell,m) = \lim_{z_0 \to 0} {z^*(t_0,{\hat y}(\tilde y(t_0,{\tilde
z_1},\mu),{\tilde z_1},\mu),z_0,\mu)\over
z_0}, 
\end{equation}
where ${\tilde z_1}$ is given in (\ref{E6.3}). From \lemref{L4.1}, we have
$z^*(\cdots) = \Phi (t_0,z_0,\mu) + z^{\$}(\cdots)$, where $\cdots$ represents 
the variables from the r.h.s. of (\ref{E6.4}), and $|z^{\$}| + |Dz^{\$}| 
= O(e^{-\alpha_1t_0})$. It follows that
$$
L_1(t_0,\ell,m) = {\partial \Phi (t_0,0,\mu)\over \partial z_0} + O(e^{-\alpha
_1t_0}),\quad \alpha_1>0.
$$
Since $\Phi(t,z_0,\mu)$ satisfies the equation $z^\prime = mz - \hat cz^3 +$ 
h.o.t., see (\ref{E4.2}),
we have ${\partial\over\partial z_0}\Phi(t_0,0,\mu)= e^{mt_0}$.  Therefore,
\begin{equation}\label{E6.5}
L_1(t_0,\ell,m) = e^{mt_0} + O(e^{-\alpha _1t_0}).
\end{equation}

Recall the monotonicity of $\Phi(t_0,z_0,\mu)/z_0,\ z_0\neq 0$, proved in 
\lemref{L4.2}.  Using the smallness of $z^\$= z^*(\ldots)-\Phi(t_0,z_0,\mu)$, 
we have the following results.

\begin{lem}\label{L6.2}  
Let $\pm z_E$ be the nonzero equilibria of (\ref{E4.2}) if $m > 0$. Let $\mu_0=
(\ell_0,m_0,k_\infty)$ where $\ell_0\in \tilde I$ and $m_0=0$.
For each $0 < \eta < 1$ there exist $\epsilon = \epsilon (\eta) >
0$ and $\bar m>0$ such that if $0 < |z_0| < \epsilon,\ |\mu-\mu_0| < 
\epsilon,\ t_0 > {\scriptstyle  {1\over \epsilon}}$, and 
$m < {\overline m}$, then we have the following result.
\begin{equation}\label{E6.6}
{d\over dz_0} [z^*(t_0,{\hat y}({\tilde y}(t_0,{\tilde z_1},\mu),{\tilde
z_1},\mu), z_0,\mu)/z_0]\qquad\qquad \qquad \qquad
\end{equation}
$$
= \left\{ \begin{array}{rlr}
{C_5(z^2_0 - \Phi^2)\over z_E^2 - z^2_0}{\Phi \over z^2_0},
&\mbox{ if } m > 0,\; z^2_E \neq z^2_0,\; z^2_0 \leq (1-\eta)\Phi^2\\
& \mbox{or } |z^2_E - \Phi^2| \leq (1-\eta)|z_E^2 - z^2_0|;
\quad &\mbox {(\ref{E6.6}a)}\\
C_6(e^{-m^\prime t_0} - 1 ) / z_0,&\mbox{ if } m > 0,\; z^2_0 = z^2_E\mbox{
and } e^{ -m^\prime t_0 } \leq 1 - \eta, \\
& \mbox{where }-m^\prime = \lim _{z \to z_E} {\partial h\over \partial z}
(\quad ); &\mbox {(\ref{E6.6}b)}\\ \displaystyle
{C_7(z^2_0 - \Phi^2)\over m-\hat cz^2_0 + z^4_0 h_1(z^0,\mu)}{\Phi \over
z^2_0},&\mbox{ if }m \leq 0, \mbox{ and } \Phi^2 \leq (1-\eta)z^2_0.
 &\mbox {(\ref{E6.6}c)}\end{array}\right.
$$
Here $C_5, C_6 > {1\over 2}$ and $C_7 > {\hat c\over 2}$ are functions of
$(t_0,\ell,m,z_0)$.  In all the three cases $\mbox { sign }
 \left \{  {\partial \over \partial z_0} ({z^*\over z_0})\right \} = - \mbox {
sign } \{z_0\}$.
\end{lem}

\bg{pf} Assume that $\bar m$ is small so that $|z_E| < \epsilon /2$. 
$$
\begin{array}{lll}
\displaystyle |{d\over dz_0} \left ({z^{\$} \over z_0}\right )| &=&
\displaystyle {|z^{\$} - (d z^{\$}/dz_0)z_0| \over |z^2_0|}\\
&\leq &\displaystyle {1\over |z^2_0|} \left \{ |z^{\$}(z_0) - {d\over dz_0} 
z^{\$}(0) \cdot z_0| + |{d\over dz_0} z^{\$}(0) \cdot z_0 - {d\over dz_0}z^{\$}
(z_0)\cdot z_0|\right \}\\ 
&\leq  &C\displaystyle \sup \| {d^2 z^{\$}\over dz^2_0}\| \cdot |z_0|.
\end{array}
$$
Here the fact that $|z_0|<\epsilon$ is used.  
We now use \lemref{L4.1}. We may assume that $\beta$ in that lemma 
 is arbitrary small at the cost of selecting
a smaller $\epsilon$. Let $0<\alpha<\alpha_0-3\beta$, we have,
$$
|{d\over dz_0}({z^{\$}\over z_0})| \leq C_8|z_0|e^{-\alpha t_0}.
$$

\begin{itemize}
\item [(i)]  Suppose $m > 0,\ z^2_E \neq z^2_0$ and $z^2_0 \leq
(1-\eta)\Phi^2$. Let $C_2$ be the constant as in \end{itemize}
\lemref{L4.2}, case (ii). Then

$$
\left |{C_2(z^2_0 - \Phi^2)\over z_E^2 - z^2_0}{\Phi \over z^2_0}\right | \geq 
\left | {C_2\eta \Phi \over z_E^2 - z^2_0}\right | \geq  
3 C_8|z_0| e^{-\alpha t_0},
$$
if $t_0 > {1\over \epsilon}$ and $\epsilon > 0$ are sufficiently small.  Here we
have used the facts  $\Phi^2 \geq z^2_0$ and $|z_E^2 - z^2_0|$ is  small.  
Therefore $|{d\over dz_0}({z^{\$}\over z_0})|<
\frac{1}{3}|{d\over dz_0}({\Phi\over z_0})|$. Since $C_2\approx 1$, $C_5>
\frac{1}{2}$.
Therefore, (\ref{E6.6}a) follows from \lemref{L4.2}, case (ii).

Suppose now $m > 0$, $z_E^2 \neq z^2_0$ and $|z^2_E - \Phi^2| 
\leq (1-\eta)|z^2_E - z^2_0|$.  Then
\begin{equation}\label{E6.7a}
\frac{|z_0^2 - z^2_E + z^2_E - \Phi^2|}{|z_E^2 - z^2_0|} \geq 1-(1-\eta)=\eta.
\end{equation}
$$
\begin{array}{lll}
\displaystyle {C_2|z_0^2 - \Phi^2|\over |z_E^2 - z^2_0|}
{|\Phi| \over z^2_0}&\geq &\displaystyle C_2\eta {|\Phi| \over z^2_0}\\
& \geq &\displaystyle 3C_8|z_0| e^{-\alpha t_0}
\end{array}
$$
The last inequality is based on the fact $|\Phi| \geq |z_0|$ and $|z_0$ is
small. Based on a similar argument, (\ref{E6.6}a) follows from \lemref{L4.2}, 
case (ii).

\begin{itemize} 
\item [(ii)]  If $m > 0,\ z^2_0 = z^2_E$ and $e^{-m^\prime t_0} \leq 1 - \eta$,
then
\end{itemize}
$$
|(e^{-m^\prime t_0} - 1) / z_0 | \geq {\eta / |z_0|} \geq
3C_8|z_0|e^{-\alpha t_0}
$$
if $|z_0| < \epsilon$ and $t_0 > {1\over \epsilon} ,\ \epsilon > 0$ is
sufficiently small. As before, (\ref{E6.6}b) then follows from \lemref{L4.2}, 
(iii).

\begin{itemize}
\item [(iii)]  If $m \leq 0$ and $\Phi^2 \leq (1-\eta)z^2_0$, then
\end{itemize}
$$
\begin{array} {ll}
&\displaystyle \big|{C_1(z^2_0 - \Phi^2)\over m-\hat cz^2_0 + z^4_0 h_1(z_0,\mu)}
{\Phi \over z^2_0}\big|\\
&\displaystyle \geq \big|{ C_1\eta\Phi\over m-\hat cz^2_0+z^4_0h_1(z_0,\mu)}\big|\\
 &\geq 3C_8|z_0|e^{-\alpha t_0},
\end{array}
$$
When deriving the last inequality, we assume $\epsilon$ is sufficiently small
so that $\sup_z\{h(z)\} \\ =\tilde m$
with $\tilde m<\alpha$. Then $|\Phi| \geq |z_0|e^{-\tilde mt_0}$ by the Gronwall
inequality. Therefore, if $t_0$ is sufficiently large and 
$m-\hat cz^2_0 + z^4_0 h_1(z_0,\mu)$ is small, the last inequality holds. 
(\ref{E6.6}c) then follows from \lemref{L4.2}, case (i).

The proof of \lemref{L6.2} has been completed.
\end{pf}

\begin{lem}\label{L6.3}  
Under the same conditions of \lemref{L6.2}, in any of
the three cases, (\ref{E6.6}a), (\ref{E6.6}b) or (\ref{E6.6}c), if we choose 
smaller $\epsilon > 0$, we have, (see (\ref{E6.3}) for $\tilde z_1$),
$$
{\partial \over \partial z_0}\left({z^*\over \tilde z_1}\right) \left\{\begin{array}{ll}
> 0,&\quad \mbox{if }  \tilde z_1 < 0,\\
< 0,&\quad \mbox{if }  \tilde z_1 > 0.
\end{array}\right.$$
\end{lem}

\bg{pf} Observe that
$\displaystyle{\partial \over \partial z_0}
\left ( {z^*\over \tilde z_1}\right ) = {\partial \over \partial z_0}
\left ( {z^* \over z_0} {z_0\over \tilde z_1}\right )
= {z_0\over \tilde z_1} {\partial \over \partial z_0} \left ( {z^*\over
z_0}\right ) + {z^*\over z_0} {\partial \over \partial z_0} \left (
{z_0\over \tilde z_1}\right ).$
Since $\tilde z_1$ is an odd function of $z_0$, we have
$$
|{\partial \over \partial z_0} \left (  {z_0\over \tilde z_1}\right ) | = |
(\tilde z_1 - {\partial \tilde z_1 \over \partial z_0} z_0)/\tilde z^2_1
| \leq C_9 |z^3_0|/\tilde z^2_1.
$$
We first show that $\displaystyle|{\partial \over
\partial z_0} \left ( {z^* \over z_0}\right ) | > 2C_9 |{z_0z^*\over
\tilde z_1}|$, since this will imply that the sign of 
${\partial \over \partial z_0}\left({z^*\over \tilde z_1}\right)$ is determined
by the sign of ${z_0 \over \tilde z_1} {\partial
\over \partial z_0} \left ( {z ^*\over z_0}\right )$.
Based on $|\Phi| \geq |z_0|e^{-\tilde mt_0}$ and $|z^{\$} | \leq C |z_0|
e^{-\alpha _1t_0}$, we have for large $t_0$, $|z^*| < 2 |\Phi|$.  We then need
to show
\begin{equation}\label{E6.9}
\left|{\partial \over \partial z_0} \left ( {z^*\over z_0}\right )\right| >
C_{10}|\Phi|,
\end{equation}
where $C_{10} = 4C_9C_{11}$ with $C_{11} = \sup |{z_0\over \tilde z_1}|$.

\begin{itemize}
\item [(i)] If $m > 0$, $z^2_E \neq z^2_0$ and $z^2_0  \leq (1-\eta )\Phi^2$,
then from (\ref{E6.6}a),
\end{itemize}
$$
\left|{\partial \over \partial z_0} \left ( {z^*\over z_0}\right)\right| >
\frac{1}{2}
{|\eta \Phi |\over |z^2_E - z^2_0|} > C_{10} |\Phi|,
$$
provided that $|z^2_E - z^2_0| < {\eta \over 2C_{10}}$, which can be achieved
by choosing  smaller $\epsilon $.
\begin{itemize}
\item [(ii)]  If $m > 0$, $z^2_E \neq z^2_0$ and $|z^2_E  - \Phi^2| \leq
(1-\eta)|z^2_E - \Phi^2|$, then from (\ref{E6.6}a) and (\ref{E6.7a}),
\end{itemize}
$$
\left|{\partial \over \partial z_0}\left({z^*\over z_0}\right)\right|>{|\eta
\Phi|\over 2z^2_0}  > C_{10}|\Phi|,
$$
provided that $z^2_0 < {\eta \over 2C_{10}}$, which is valid if $\epsilon$
is sufficiently small.

\begin{itemize}
\item[(iii)]  If $m>0,\,z_0^2=z_E^2$ and $e^{-m't_o}\leq 1-\eta$, then
from (\ref{E6.6}b),
\end{itemize}
$$\left|{\partial\over\partial z_0}\left({z^*\over z_0}\right)\right|\geq
{\eta\over 2|z_0|} \geq C_{10}|\Phi|.
$$
Here we need $\displaystyle |z_0\Phi| < {\eta\over 2C_{10}}$, which is valid
if $\epsilon$ is small.

\begin{itemize}
\item [(iv)]  If $m \leq 0$ and $\Phi^2 \leq (1-\eta)z^2_0$, then from 
(\ref{E6.6}c),
\end{itemize}
$$
\left|{\partial \over \partial z_0}\left({z^*\over z_0}\right)\right|\geq{|\eta
\Phi|\over 2|m-\hat  cz^2_0 + z^4_0h_1(z_0,\mu)|} \geq C_{10} |\Phi|,
$$
provided that $|m - \hat cz^2_0 + \ldots \mid \leq {\eta \over 2C_{10}}$,
that is  valid if $\epsilon $ is small.

In all the cases, the sign of ${\partial \over \partial z_0} \left ( {z^*\over
\tilde z_1}\right )$ agrees with that of ${z_0 \over \tilde z_1} {\partial
\over \partial z_0} \left ( {z ^*\over z_0}\right )$.  From \lemref{L6.2}
$$
\mbox { sign } \left \{ {\partial \over \partial z_0} \left ( {z^*\over
z_0}\right ) \right \} = - \mbox { sign } \{z_0\}.
$$
Therefore $\mbox { sign} \left \{ {\partial \over \partial z_0}\left ( {z^*
\over \tilde z_1}\right )\right \} = - \mbox { sign} \left \{\tilde z_1\right
\}$\hfill  
\end{pf}

\begin{cor}\label{C6.4} Under the conditions of \lemref{L6.2}, we have 
$\displaystyle |{z^*\over z_1}|$ is a decreasing function of $|z_1|$.
\end{cor}

Recall that $t_1$ is fixed and $T=t_0+t_1$ depends solely on $t_0$. Let
\bg{equation}\label{E6.9a}
\bg{aligned}
L(T,\ell,m) &= L_1(t_0,\ell,m)e^{m t_1},\\
r(T,\ell,m) &= r_1(t_0,\ell,m)e^{-m t_1}.
\end{aligned}
\end{equation}
We shall prove that with such functions $L$ and $r$, \thmref{T3.4} is valid.

If no arguments are given, for notational simplicity, $\Phi$ means
$\Phi(t_0,z_0,\mu)$, $z_0$ is defined in (\ref{E6.1}), $z^* = G_2(t_0,\ell,m,z_1)$ is
defined in (\ref{E5.10}), $z_1 = \tilde z_1$ is defined in (\ref{E6.3}), $z^{\$}$ means $z^{\$}
(t_0,\hat y(\tilde y(t_0,\tilde z_1,\mu), \tilde z_1,\mu),z^0,\mu) = z^* -
\Phi$.  Assumptions following a case number are valid until a new case is
encountered.

\bg{pf*}{\bf Proof of \thmref{T3.4}}  From (\ref{E6.9a}), if $m=0$, 
$$r(\infty,\ell_0,0)=r_1(\infty,\ell_0,0).
$$
Thus, \thmref{T3.4}, 1) follows from \lemref{L6.1}.

From (\ref{E6.9a}) again, it is easy to see that 2) follows from (\ref{E6.5}).

We now prove \thmref{T3.4}, 3).  From (\ref{E6.4}) and \lemref{L4.1},
\begin{eqnarray*}
 L_1(t_0,\ell,m)&=&{\partial \Phi(t_0,0,\mu)\over\partial z_0}+
\left .{\partial z^{\$} \over\partial z_0}\right |_{z_o=0},\\
\left | {\partial z^\$ \over \partial z_0} \right | + 
\left | {\partial^2 z^{\$}\over \partial z_0  \partial m}\right | &\leq &
Ce^{(-\alpha _1+2\beta) t_0},\quad \alpha_1>2\beta>0,\\ 
{\partial \Phi (t_0,0,\mu)\over \partial z_0}
&=& e^{mt_0},
\end{eqnarray*}
we have
$$
{\partial L_1(t_0,\ell,m)\over \partial m} = t_0 e^{mt_0} + O(e^{(-\alpha _1
+ 2\beta) t_0}).
$$
Now that ${\partial \over \partial m}(L_1r_1) = {\partial L_1\over \partial m} 
r_1 + L_1{\partial r_1\over \partial m}$ and 
$|{\partial r_1\over \partial m}| \leq C$, we
have ${\partial \over \partial m}(L_1r_1) = t_0e^{mt_0} \cdot r_1(t_0,\ell,m) +
O(e^{mt_0})$.  Let ${\cal O} = \{|\ell - \ell _0| <
\delta, |m| < \delta \}$ and let $t_0>1/\delta$. If $\delta > 0$ is 
sufficiently small, we
have $r_1(t_0,\ell,m) = C_1c^*(\ell _0)$ where $C_1 > {\frac{1}{2}}$.  Thus, 
$$
\mbox { sign } \left \{ {\partial \over \partial m} (L_1r_1)\right \} = \mbox
{ sign } \left \{ c^*(\ell _0)\right \}.
$$
The assertion in 3) follows by observing that $L\cdot r=L_1 \cdot r_1$.

Consider $c^*(\ell _0) > 0$ and look for SN simple
periodic solutions first.  We only need to solve (\ref{E5.11}) for $z_1 > 0$ 
since
$G_2(t_0,\ell,m,z_1)$ is odd in $z_1$.  If $z_1$ is sufficiently small and
$t_0$ is sufficiently large, we have $r_1(t_0,\ell,m) >
0$ and $z_0/z_1 > 0$.  Thus we need to consider $z_0 > 0$ only. We now look
for a solution $z_1\in(0,\zeta)$ with the corresponding $z_0\in(0,\epsilon)$
where $(-\epsilon,\epsilon)$ is the coordinate chart in the $z$-axis for 
$W^c_{loc}(E)$. 

Since $z_0=0$  if $z_1=0$ and $z_0$ depends continuously on $z_1$, we can choose
 a small constant $\zeta  > 0$ so that $z_1 = \zeta $ implies
$z_0 < \epsilon$. If $\delta$, which defines the set ${\cal O}$, is small, then
 either $z_E < \zeta /2,\,m>0$ or $z_E$ does
not exist $(m\leq 0)$.  We can choose $\overline t > 0$
so large that $t_0>
\overline t$ implies that $0<\Phi < 3\zeta /4$, and $|z^{\$}|<\zeta /4$. 
The first estimate is based on $\Phi\to z_E<\zeta/2$ or $0$ as
$t_0\to \infty$. The second estimate uses \lemref{L4.1}. Therefore 
$G_2(t_0,\ell,m,\zeta )=z^*=\Phi+z^\$ < \zeta $.

Suppose now $L_1(t_0,\ell,m)r_1(t_0,\ell,m)>1$.  Then
$$
\lim _{z_0\to 0^+} {z^*\over z^0} {z^0 \over z^1} > 1.
$$
From this, there exists a small $0 < z_1 < \zeta $ such that $G_2(t_0,\ell,m,z_1) > z_1$.
Thus, there exists at least one solution $0<z_1<\zeta $ for (\ref{E5.11}) if
$Lr=L_1r_1>1$.

In the rest of the proof, we discuss the uniqueness or nonexistence of SN simple
periodic solutions. Please refer to Figure 4.1 for the flow on $W^c_{loc}(E)$.

\noindent {\bf Case (i)}: $0 < c^*(\ell _0) < 1$. For any $0<\eta_1<1$, by
choosing smaller $\delta$ and $\epsilon$,
\begin{equation}\label{E6.10}
0 < \eta _1 \leq z^2_0/z^2_1 \leq 1-\eta _1. 
\end{equation}

Let $m \leq 0$ and $z_0 > 0$.  Then $0 < \Phi \leq z_0$ and $|z^{\$}| < z_0
e^{-\alpha _1t_0}$.  If $t_0$ is sufficiently large, from \lemref{L4.1}, we have
$(z^*)^2/z_0^2 < 1/(1-\eta _1)$.  Combine this with (\ref{E6.10}), we have
$(z^*)^2 < z^2_1$. Therefore, there is no solution for (\ref{E5.11}).

Let $m > 0$ and $z_0 > z_E$.  Similar to the previous case, we find no solution
for (\ref{E5.11}).

For $m > 0$, define
$$
z_M = \sup\{ z_0\mid z^2_0 \leq (1-\eta)\Phi^2,0 \leq z_0 \leq z_E\}
$$
for some $0 < \eta < \eta_1$.  Clearly $0 \leq z_M < z_E$.

Let $m > 0$ and $z_0 \in (z_M,z_E)$.  We have $z^2_0 > (1-\eta)\Phi^2$.  By
choosing a smaller $\delta $ we have $z^2_0 > (1-\eta_1)(z^*)^2$.  From
(\ref{E6.10}), we have $(z^*)^2<z^2_1$.  There is no solution for (\ref{E5.11}) 
in this case.

Let $m > 0$, and $z_0 \in (0,z_M]$ if $z_M > 0$. At $z_0 = z_M$ we have
${\partial \over \partial z_0} \left ( {\Phi \over z_0}\right ) < 0$,  based on
\lemref{L4.2}.  We infer that $z^2_0 \leq (1-\eta)\Phi^2$ for all $z_0 \in
(0,z_M]$.  \lemref{L6.3} implies that $(z^*/z_1)$ is strictly decreasing in
that interval.  This
proves the uniqueness of solutions of (\ref{E5.11}) in this case.

Let $m > 0$ and $1 \geq L_1(t_0,\ell,m)r_1(t_0,\ell,m) = \lim_{z_1 \to 0^+}
(z^*/z_1)$.  In the case $z_0 \in (0,z_M]$, arguing as in the previous case,
we have $z^*/z_1 < 1$.  This proves
the nonexistence of solutions of (\ref{E5.11}) if $Lr \leq 1$.

\noindent {\bf Case (ii)}:  $C^*(\ell _0) > 1$. For any $0<\eta_1<1$, by  
choosing smaller $\epsilon$ and $\delta$,
\begin{equation}\label{E6.11}
z^2_1 \leq (1-\eta _1) z^2_0.
\end{equation}

Let $m > 0$ and $0 < z_0 \leq z_E$.  Since $\Phi \geq z_0$ and $|z^{\$}| \leq
z_0e^{-\alpha_1t_0}$, we have $(z^*)^2 > (1-\eta_1) z^2_0$ if $t_0$ is large
enough.  From (\ref{E6.11}), there is no solution to (\ref{E5.11}) in this case.

For $m > 0$, define
$$
z_m = \inf \{z_0\mid \Phi^2 \leq (1-\eta) z^2_0 \mbox { and } z_E < z_0 \leq
\epsilon \}
$$
for some $0 < \eta < \eta _1$. From the phase diagram, cf. Figure 4.1,  
$z_E < z_m < \epsilon $ if $\eta$ is sufficiently small.

Let $m > 0$ and $z_0 \in (z_E,z_m)$.  Then $\Phi ^2 > (1-\eta)z^2_0$.  We can
have $(z^*)^2 > (1-\eta _1)z^2_0$ if we choose $t_0$ large enough.  Then $z^*
> z_1$ based on (\ref{E6.11}). Equation (\ref{E5.11}) has no solution in this 
case.

Let $m > 0$ and $z_0 = z_m$.  Then $\Phi ^2 \leq (1-\eta)z^2_0$.   This
implies that $|\Phi^2 - z^2_E|\leq \linebreak (1-\eta) \mid z^2_0 - z^2_E|$.
We then can show that $\Phi^2/z_0^2$ is decreasing for $z_0 \in [z_m,\epsilon]$.
Thus $\Phi ^2 \leq (1-\eta)z^2_0$ for $z_0\in [z_m,\epsilon]$.
From \lemref{L6.3}, $z^*/z_1$  is strictly decreasing.  The solution to 
(\ref{E5.11}) is either
unique or does not exist.  We show $L_1(t_0,\ell,m)r_1(t_0,\ell,m) > 1$ in this
case, so that the nonexistence becomes  impossible.  In fact,
if $\epsilon  > 0$ is small, then $r_1(t_0,\ell,m) > 1 + \eta _2$ for
some $\eta _2 > 0$, due to $c^*(\ell_0)>1$.  Because
$m > 0$ and $L_1(t_0,\ell,m) = e^{mt_0} + O(e^{-\alpha_1 t_0})$,  let $t_0$ be
sufficiently large,  we have $L_1 > (1+\eta _2)^{-1}$.  Therefore $L_1r_1 > 1$.

For $m \leq 0$, define
$$
z^m = \mbox { inf } \{z_0 \mid \Phi ^2 \leq (1-\eta) z^2_0,\ 0 \leq z_0 \leq
\epsilon \}
$$
for some $\eta < \eta _1$ and $\eta<1/4$.  Clearly $0 \leq z^m < \epsilon $ if
$\eta$ is sufficiently small.

Let $m \leq 0$ and $z_0 \in (0,z^m)$ if $z^m > 0$.  Then $\Phi ^2 >
(1-\eta)z^2_0$.  We can make $(z^*)^2 > (1-\eta _1)z^2_0$ by choosing
$\epsilon < 0$ smaller.  Therefore $z_1 < z^*$ by (\ref{E6.11}).  There is no
solution to (\ref{E5.11}).

Let $m \leq 0$ and $z_0 = z^m$.  Then $\Phi^2\leq (1-\eta)z^2_0$.  By 
\lemref{L4.2},
$\Phi ^2/z^2_0$ is decreasing for $z_0 \in [z^m, \epsilon )$.  Thus $\Phi\leq
(1-\eta)z^2_0$ for $z_0\in[z^m,\epsilon]$. Then $z^*/z_1$ is
strictly decreasing by \lemref{L6.3}.  There is either no solution or the solution is unique to
(\ref{E5.11}) when $z_0 \in [z^m, \epsilon )$.

Let $m \leq 0$ and $0\leq L_1(t_0,\ell,m) r_1(t_0,\ell,m) \leq 1$.  By 
(\ref{E6.11}),
$r_1^2\geq (1-\eta _1)^{-1}$.  Thus, $L_1^2 < 1-\eta _1$.  If we choose
$\epsilon > 0$ small,
we have $(z^*)^2 / z^2_0 < 1 - \eta _2$ for some $0 < \eta _2 < \eta _1$.  And
also $\Phi ^2 / z_0^2 < 1 - \eta$ for some $0 < \eta < \eta _2$ in the interval
$[z^m,\epsilon]$.  Thus
$z^*/z_1$ is strictly decreasing.  Since $\lim _{z_1 \to 0} z^*/z_1 \leq 1$,
there is no solution to (\ref{E5.11}) in this case.

\noindent {\bf Case (iii)}:  $C^*(\ell _0) = 1$. For any $0< \eta<1$, we can 
choose a smaller
$\epsilon$ so that
\begin{equation}\label{E6.12}
1 - \eta < (z_0/z_1) < 1 + \eta .
\end{equation}
If $\epsilon$ is small, then $r_1(t_0,\ell,m)<1+\eta$.
Let $L_1(t_0,\ell,m)r_1(t_0,\ell,m) > 1 + \delta$ for some $\delta > \eta$.  Then
$$
L_1 > {1+\delta \over 1 + \eta} \geq 1 + \eta _1
$$
for some $0 < \eta _1 < \delta$.  From $L_1(t_0,\ell,m) = e^{mt_0}+
O(e^{-\alpha _1 t_0})$,
if $t _0$ is large enough, we have $m > 0$, and $mt_0 > \epsilon _1$ for
some $\epsilon _1 > 0$.  From \lemref{L4.3}, case (a), we have $z^2_0 \leq (1-\eta
_3)\Phi ^2$, or $|\Phi^2 - z_E^2|\leq (1-\eta _3)|z^2_0-z^2_E|$ or $e^{mt_0} -
1 \geq \eta _3$ for some $\eta _3 > 0$.  Therefore $z^*/z_1$ is strictly
decreasing and the solution to (\ref{E5.11}) is unique.

Let $L_1(t_0,\ell,m)r_1(t_0,\ell,m) < 1-\delta$, for some $\delta > \eta$.  By a
similar argument, $m t_0 < -\epsilon _1$ for some $\epsilon _1 > 0$.  Also we
must have $m < 0$.  From \lemref{L4.3}, case (b), we then have $\Phi ^2 \leq
(1-\eta _3)z^2_0$ for some $ \eta _3 > 0$.
Thus $z^*/z_1$ is strictly decreasing.  There is no solution to
(\ref{E5.11}) since $\lim _{z_1 \to 0^+} z^*/z_1 = L_1 \cdot r_1 < 1 - \delta$.

We have completed the discussion for the case $C^*(\ell _0) > 0$.

Consider $C^*(\ell _0) < 0$ and symmetric double periodic SN solutions next.  We need to
solve (\ref{E5.11})${}^\prime$.  We can divide the case into three subcases---case
(iv): $-1< C^*(\ell _0) < 0$, case (v): $C^*(\ell _0) < -1$ and  case (vi):
$C^*(\ell _0) = -1$.  They are analogous to cases (i), (ii) and (iii)
respectively.  The proofs are  similar to the previous cases and will not
be rendered here.

This completes the proof of \thmref{T3.4}.
\end{pf*}

\bg{pf*}{\bf Proof of \thmref{T3.5}}  From our assumption, $c^*(\ell_0)=r_1(\infty,\ell_0,m_0) = 0$ and \newline
${\partial \over
\partial \ell}  r_1(\infty, \ell _0,m_0) \neq 0$.  Using the implicit function
theorem we can find a unique $C^1$ function $\ell = \ell ^*(m)$ so that
$r(\infty, \ell ^*(m) , m) = 0, \ |m| < \delta$.

Since $|\tilde y(t_0,z_1,\mu)\mid < Ce^{-\alpha _1t_0}$ and
$|k^*(t_0,\ell,m,z_1) - k^*(\infty,\ell,m,z_1)\mid  < Ce^{-\alpha _1t_0}$, we
have $r_1(t_0,\ell ^*(m),m) = O(e^{-\alpha_1t_0})$. Please refer to (\ref{E6.1}),
(\ref{E6.2}) for the definitions of $z_0$ and $r_1$.

Since $z_0$ is an odd function of $z_1$, for some $\bar C>0$, we have
$$
\left | {z_0 \over z_1}\right | \leq \overline C (e^{-\alpha _1t_0} + z^2_1 +
|\ell - \ell ^*(m)|).
$$
Since $\Phi (t_0,z_0,\mu)$ satisfies the equation $z^\prime = mz - \hat
cz^3+\cdots$, we have $|{\Phi \over z_0}| \leq e^{mt_0}$ if $|z_0| < \epsilon$
and $|\Phi| < \epsilon$.  Thus, from \lemref{L4.1},
$$
\left | {z^*\over z_0}\right | \leq e^{mt_0} + Ce^{-\alpha _1t_0} \leq
2e^{mt_0},
$$
if $ t_0 > \overline t$ is sufficiently large.  We now choose $\delta (T)
= Ce^{-mT}$ where $C$ is a small constant.  If $|\ell -\ell ^*(m)|<\delta
(T)$ and $z_1<(\delta (T))^{1/2}$, then $\overline C(e^{-\alpha _1t_0} +
z^2_1 + |\ell -\ell ^*(m)|) < {\frac{1}{2}} e^{-mt_0}$.  Therefore,
 $\left | {z^*\over z_0}\right |\  \left | {z_0\over z_1}\right | < 1$.
The bifurcation equation (\ref{E5.11}) or (\ref{E5.11})${}^\prime$ has no solution in this case.
\end{pf*}

%% file: sect7.tex
\section {Numerical test on a predator-prey model}

The following predator-prey
model was proposed by Freedman and Wolkowicz [13] to describe group defense of
prey against predatation.
\bg{equation}\label{E7.1}
\begin{aligned}
{\dot u} & = 2u(1-{u\over k}) - 9vp(u),\\
{\dot v} & = v(-\gamma + 11.3 p(u)),\\
u & = \mbox { prey, } \quad v\quad  = \quad \mbox {predator}.
\end{aligned}
\end{equation}
where $p(u) = u/(u^2 + 3.35u + 13.5)$ represents the interaction between prey
and predator.  For a large range of $(\gamma,k)$, (\ref{E7.1}) has
two interior equilibria $(\bar u_0,\bar v_0)$ and $(\bar{\bar u}_0,\bar{\bar
v}_0)$, with   $\bar {\bar u}_0 < \bar u_0$.
Here $p(\bar u_0) = p(\bar{\bar u}_0) =
{\gamma \over 11.3}$, while $\bar v_0,\,\bar{\bar v}_0$ can
be solved from the first equation of (\ref{E7.1}).  We are interested in the
equilibrium $(\bar u_0,\bar v_0)$, which is hyperbolic, and shall be denoted
by $E= E(\gamma,k)$.

Let the Jacobian matrix at $E$ be $\left ( \begin{array}{cc}  -a & -b\\ -c &
-d\end{array} \right )$.  
It is shown in [25] that $a >0,\,b>0,\,c>0$ and $d=0$.  Thus, Hypothesis $H_3$) 
in $\S 3$ is satisfied.

Freedman and Wolkowicz  have discovered a curve $\cal{S}\subset{\Bbb R}^2$
such that if $(\gamma,k)\in\cal{S}$, then (\ref{E7.1}) possesses a homoclinic
solution $q(t)$ asymptotic to the equilibrium $E(\gamma,k)$. Numerical 
computation shows
that the curve $\cal S$ can be
parameterized by $\bar u_0$ and is plotted in Figure 7.1.  For each
$(\gamma,k)\in\cal{S}$, let $\gamma$ be fixed and let $k$ vary. Then the
homoclinic solution breaks. The derivative of the gap between $W^u(E)$ and
$W^s(E)$ with respect to $k$ can be evaluated by the  Melnikov
integral $M_\gamma(k)$, as in $H_4$). The Melnikov integral has been computed
numerically, and the result
is plotted in Figure 7.2.  Evidently, $M>0$ for all the values considered.
Thus, Hypotheses $H_2$) and $H_4$) in $\S 3$ are satisfied for those parameter
values.
\vfill
\vskip1.2in
\centerline{\bf Fig. 7.1}
\vskip1.2in
\centerline{\bf Fig. 7.2}
\vskip 0.05in

The smooth dependence of $M_\gamma(k)$ on $\bar u_0$ indicates that $M>0$ is
not a numerical artifice.

In the remaining of this section, we fix $(\bar u_0,\,\gamma,\,k)$ =
(5.49178,\,1.0,\,6.87433).  After adding diffusions $(d_1u_{\xi\xi},\
d_2v_{\xi\xi})$, we consider a system of PDEs in the domain $0 < \xi < 1$ with
Neumann boundary conditions, cf.\ (\ref{E1.2}).  Let $\Gamma$ be the curve in
$(d_1,d_2)$-plane on which (1.2) has a zero eigenvalue with associated
eigenvectors in ${\cal X}_1$.  Since $bc > 0$ and $a > 0$, $\Gamma$ is
depicted in Figure 3.1, Case 1.

For $(d_1,d_2) \in \Gamma$, we now compute $W^c_{loc} (E)$ and the flow on it,
up to $O(\rho^3)$ where $\rho = |u-\bar u_0| + |v-\bar v_0|$.   Since the
boundary conditions are of the Neumann
type, we will expand $(u,v)$ into Fourier cosine series.
Let $(\bar u_0+\sum^\infty_0 u_n\cos n\pi\xi,\, \bar v_0+
\sum^\infty_0 v_n \cos n \pi \xi) \in
W^c_{loc}(E)$. Let $(u_1\cos \pi\xi,\,v_1\cos\pi\xi)$, with $(u_1,v_1)
=(\eta,1)$, be the unique eigenvector corresponding to the eigenvalue
$0$, up to a constant multiple. Then $u_1 = \eta v_1 + \phi(v_1),\,
u_n = u^*_n(v_1)$ and
$v_n = v^*_n(v_1)$ for $n \neq 1$, where $\phi,\ u^*_n,\ v^*_n = O(v^2_1)$.
This is due to the
fact that $W^c_{loc}(E)$ is tangent to the zero eigenvector corresponding
to $(u_1,v_1) = (\eta,1)$.  Because of the $R$ symmetry (Theorem 2.7), we have
$\phi = O(\rho^3)$.

The Fourier coefficients $(u_n,\ v_n)$ are functions of $t$. They satisfy
\bg{equation}\label{E7.2}
\begin{array}{ll}
u'_n = - d_1n^2\pi^2u_n - au_n - bv_n + [f(u,v)]_n+O(\rho^4),\\
v'_n = - d_2n^2\pi^2v_n - cu_n+[g(u,v)]_n+O(\rho^4),
\end{array}
\end{equation}
where $f$ and $g$ are polynomials of degree $3$.
For $h\in L^2(0,1)$,
we use $[h]_n$ to denote the $n$th Fourier cosine coefficient for $h$.
Using some basic trigonometry formulas, we can rewrite $[f(u,v)]_n$ and
$[g(u,v)]_n$ in terms of $\{u_n\}_0^\infty,\, \{v_n\}_0^\infty$. Only finitely
many terms are needed here since other terms will be included in $O(\rho^4)$.

We can now use the Taylor expansion method in [2] to obtain a power series
expansion of $\phi(v_1)$ and the flow on the center manifold. The function $\phi$
has the form $\phi(v_1)=c v_1^3+O(v_1^5)$. And the flow on the center manifold
has the from

$$
v'_1 = \tilde cv^3_1 + O(\rho^4).
$$
\vskip1.2in
\centerline{\bf Fig. 7.3}
\vskip 0.05in

When $(d_1,d_2)$ moves along the curve $\Gamma$,  values of $\tilde c$ have been
computed numerically and the results are depicted in Figure 7.3, with $\hat c$
against $\pi^2 d_1$.  It  verifies
that $\tilde c < 0$ for the portion of $\Gamma$ under consideration.
Since there is a diffeomorphism between $v_1$ and $z$, Hypothesis $H_5$), in 
\S 3 has been verified numerically.
\vskip1.2in
\centerline{\bf Fig. 7.4}
\vskip 0.05in

We now compute $c^*(d_1,d_2)$ for $(d_1,d_2) \in \Gamma,\ 0 < \pi^2 d_1 < 3$.
Again,
we fix $u_0$=5.49178, $\gamma =1$ and $k$=6.87433.  Numerical results of
$c^*$ are depicted in Figure 7.4.  We have found a point $d^*_1$=0.183 such
that $c^*(d_1,d_2) < (> \mbox { or } = )\ 0$ if $d_1<(>\mbox { or } = )
\ d^*_1$. The results also show that
$\partial c^*(\ell _0)/\partial \ell \neq 0$ where
$(\ell _0,0) \in \Gamma$ corresponds to $d_1=d^*_1$.  Therefore all the
twisted, nontwisted and degenerate cases have be found in Freedman and
Wolkowicz's example.  However the case $c^* (\ell _0) \geq 1$ or $\leq -1$ has
not been found in this example. Numerical and theoretical result also indicate
that there is a point $(d_1,d_2)=(0.0093,0.0093)$ where $c^*=0$. However, the 
numerical
error near that point is too large to be trustworthy. Thus, we do not include it
in Figure 7.4.

We end this section by proving \lemref{L6.1}.
\bg{pf*}{\bf Proof of Lemma 6.1}
Recall the definition of
$\hat z$ in \S 5 and $r_1$ in (\ref{E6.1}) and (\ref{E6.2}).  We need to 
consider the $z$-th component of
${\partial \over \partial z_1} U_*(t_1,\overline x,y_1,z_1,\mu)$, with $y_1 =
0$ and $z_1 = 0$.  Let ${\partial \over \partial z_1} U_*(t) =
(x(t),y(t),z(t))$, $0 \leq t \leq t_1$.  It satisfies the linear variational
equation (3.4) and the initial conditions are $x(0) = 0, y(0) = 0,z(0) = 1$.
We now extend the solution $(x(t), y(t), z(t))$ to $t \leq 0$. Notice that we
are treating an infinite dimensional system, so the backwards extension of a
solution is not unique. Using the flat coordinates (4.1), in a neighborhood
of $0$, we write (3.4) as
\bg{equation}\label{E7.8}
\begin{array}{lll}
x^\prime&=&  A_1x+(D_xg_1)x+(D_yg_1)y+(D_zg_1)z,\\
y^\prime&=&  A_2y+D_yg_2(x_q(t),y_q(t),z_q(t),\mu)y,\\
z^\prime&=&  A_3z+(D_xg_3)x+(D_yg_3)y+(D_zg_3)z,
\end{array}
\end{equation}
where $q(t)=(x_q(t), y_q(t),z_q(t))$ in the flat coordinates. 
Here we have used the facts $y_q(t)=0,\,t\leq 0$, and $g_2(x,0,z,\mu)=0$
to simplify the second equation of (\ref{E7.8}).

First let $y(t) \equiv 0$
for $t \leq 0$, which solves the second equation. Then $(x(t), z(t))$ can be
solved uniquely from (\ref{E7.8}) backward in
time.  We now show $x(t) \equiv 0$ for $t \leq 0$.  If $(x(t),0,z(t))$ is a
solution for (\ref{E7.8}), so is $R(x(t),0,z(t)) = (x(t),0,-z(t))$.  Thus,
$(x(t),0,0)$
is a solution of (\ref{E7.8}).  Since $x(0)=0$, solving the one-dimensional
ODE for
$x(t)$ we have $x(t) \equiv 0$ for $t\leq 0$. Observe that $g_1(0,y,z,\mu) = 0$.
Thus, the first equation is valid even if $z\neq 0$.  The equation for $z(t)$
becomes
\bg{equation}\label{E7.9}
z' = A_3z+D_zg_3(x_q(t),y_q(t),z_q(t),\mu)z,
\end{equation}
with $A_3=0$. In our flat coordinates, $z_q(t) = 0$ and $y_q(t) = 0$, we have
\newline $D_zg_3(x_q(t),0,0,\mu) = D_zg_3(0,0,0,\mu) = 0$, since zero is an
eigenvalue for $(d_1,d_2) \in \Gamma$.  Here we used the
fact $g_3(x,0,z,\mu) = g_3(0,0,z,\mu)$ on $W^{cu}_{loc}$.  Thus, (\ref{E7.9})
becomes $z'=0$, and $z(t) \equiv 1$
for $t \leq 0$.  $(x(t),y(t),z(t)) = (0,0,1)$ for $t \leq 0$.

We now have $(x(0),y(0),z(0)) \in {\cal X}_1$ and shall remain in ${\cal X}_1$
for $t \geq 0$.  In particular, $x(t) = 0$ for all $t \in R$.
According to $\S 3$,
$(x(t),y(t),z(t)) \rightarrow (0,0,c^*)$ as $t \rightarrow + \infty$.  However,
because the coordinates are flat, $g_3(0,y,z,\mu) = g_3(0,0,z,\mu)$.  Also
$x_q(t) \equiv 0$ for $t \geq t_1$.  Thus ${\partial \over \partial
y}g_3(0,y_q(t),0,\mu) = 0$ for $t \geq t_1$.  Again $z(t),\,t\geq t_1$
satisfies (\ref{E7.9}) with $x_q(t)=0$ and $z_q(t)=0$.  Since
$D_zg_3(0,y_q(t),0,\mu) = D_zg_3(0,0,0,\mu) = 0$, We have $z(t)$ =
constant for $t \geq t_1$.  Thus $z(t_1) = c^*$.  This proves \lemref{L6.1}.
\end{pf*}
\vskip 0.05in

\noindent{\bf Acknowledgment.}  I would like to thank R. Silber
for teaching me Pascal programming on IBM PCs and W. McKinney for
helping  me with unix operating systems. I sincerely  thank the 
referees  for their comments that have helped to improve the presentation
of this paper. This research is partially supported
by the National Science Foundation under Grant No. NSF-DMS9002803 and
DMS9205535.

%% file: biblio.tex
\makeatletter \renewcommand{\@biblabel}[1]{\hfill#1.}\makeatother

%% file: figures.tex
\renewcommand{\baselinestretch}{1.8}

\bg{center}\textbf{FIGURE CAPTIONS}
\end{center}
\vskip 0.1in

\no{\bf Figure 1.1.} Twistedness of the homoclinic orbit is determined by
comparing $\phi(-T)$ and $\phi(T)$.
\vskip 0.05in

\no{\bf Figure 3.1.} The curve $\Gamma$ divides the first quadrant of the
$(d_1,\,d_2)$ plane into two parts. All the possibilities are listed, except for
permutations of $d_1$ and $d_2$.
\vskip 0.05in

\no{\bf Figure 3.2.} A sketch of the bifurcation diagram in the $(\ell,\,m)$
plane. SN simple or symmetric double periodic solutions occur in the shaded
areas.
\vskip 0.05in

\no{\bf Figure 3.3.} A sketch of all kinds of homoclinic, heteroclinic and
periodic solutions.
\vskip 0.05in

\no{\bf Figure 4.1.} Phase diagrams for the flow on the center manifold.
\vskip 0.05in

\no{\bf Figure 5.1.} A sketch of the inner solution $U^*$ and outer solution 
$U_*$. 
\vskip 0.05in

\no{\bf Figure 7.1.} Values of $(\gamma,\,k)$ where a homoclinic orbit to (7.1)
exits are plotted, using $u_0$ as an independent variable.
\vskip 0.05in

\no{\bf Figure 7.2.} The Melnikov integral $M_\gamma(k)$ is plotted, 
 using $u_0$ as an independent variable.
\vskip 0.05in

\no{\bf Figure 7.3.} Values of $\tilde c$ are plotted when $(d_1\,d_2)$ moves 
along $\Gamma$, which are parameterized by $d_1$.
\vskip 0.05in

\no{\bf Figure 7.4.} Values of $c^*$ is plotted when $(d_1\,d_2)$ moves 
along $\Gamma$, which is parameterized by $d_1$.
\vskip 0.05in